\documentclass[11pt,prd,groupedaddress,showpacs,doublecolumn,nofootinbib]{revtex4-1}

\usepackage{graphicx}
\usepackage{dcolumn}
\usepackage{bm}
\usepackage{amssymb}
\usepackage{amsmath}
\usepackage{epsfig}
\usepackage{hyperref}
\usepackage{hhline}
\usepackage{color}
\usepackage{multirow}
\usepackage[normalem]{ulem}
\usepackage{slashed}
\usepackage{slashed}
\usepackage{tikz}
\usetikzlibrary{snakes}
\usepackage{hyperref}
\usepackage[section]{placeins}

\definecolor{Zsug}{RGB}{0, 145, 33} 
\definecolor{Zcor}{RGB}{210, 0, 210}
\definecolor{Zque}{RGB}{0, 180, 190} 
 
\definecolor{jd}{rgb}{0.858, 0.188, 0.478}

\def\lapp{\mathrel{\rlap{\raise.5ex\hbox{$<$}}
                    {\lower.5ex\hbox{$\sim$}}}}
\def\gapp{\mathrel{\rlap{\raise.5ex\hbox{$>$}}
                    {\lower.5ex\hbox{$\sim$}}}}

{\newcommand{\lsim}{\mbox{\raisebox{-.6ex}{~$\stackrel{<}{\sim}$~}}}
{\newcommand{\gsim}{\mbox{\raisebox{-.6ex}{~$\stackrel{>}{\sim}$~}}}

\newcommand{\bmt}{\begin{pmatrix}}
\newcommand{\emt}{\end{pmatrix}}
\newcommand{\ba}{\begin{array}{c}}
\newcommand{\ea}{\end{array}}
\newcommand{\be}{\begin{equation}}
\newcommand{\ee}{\end{equation}}
\newcommand{\bea}{\begin{eqnarray}}
\newcommand{\eea}{\end{eqnarray}}

\newcommand{\bi}{\begin{itemize}}
\newcommand{\ei}{\end{itemize}}

\newcommand{\baz}{\begin{array}{cc}}

\newcommand{\mathsym}[1]{{}}

\newcommand{\bt}{\begin{tabular}}
\newcommand{\et}{\end{tabular}}

\newcommand{\benu}{\begin{enumerate}}
\newcommand{\eenu}{\end{enumerate}}

\newcommand{\bav}{\begin{array}{cccc}}

\begin{document}

\begin{flushright}
IP-BBSR-2019-12
\end{flushright} 
  
\renewcommand*{\thefootnote}{\fnsymbol{footnote}}

\begin{center}
 {\Large\bf A Minimal Model of Torsion Mediated Dark Matter}
 \\
 \vskip .5cm
 {
  Basabendu Barman$^{a,}$\footnote{bb1988@iitg.ac.in},
 Tapobroto Bhanja$^{a,}$\footnote{tapobroto.bhanja@iitg.ac.in},
 Debottam Das$^{b,}$\footnote{debottam@iopb.res.in},
 Debaprasad Maity$^{a,}$\footnote{debu@iitg.res.in},
 }\\[3mm]
 {\it{
 $^a$ Department of Physics, Indian Institute of Technology Guwahati, Assam 781039, India \\
 $^b$Institute of Physics, Bhubhaneswar, Odhisha- 751005, India  \& Homi Bhabha National Institute,
Training School Complex, Anushakti Nagar, Mumbai 400085, India}\\
 }
 \end{center}
\vspace{1cm}

\begin{center}
 {\bf{Abstract}}
\end{center}

\begin{abstract}
  We present a minimal model of fermionic dark matter (DM), where a singlet Dirac fermion can interact with the Standard Model (SM) particles via the torsion field of gravitational origin. In general, torsion can be realized as an antisymmetric part of the affine connection associated with the spacetime diffeomorphism symmetry and thus can be thought of as a massive axial vector field. Because of its gravitational origin, the torsion field couples to all the fermion fields including the DM with equal strength, which makes the model quite predictive. The DM is naturally stable without any imposition of ad-hoc symmetry {\it e.g.,} $\mathcal{Z}_2$. Apart from producing the correct thermal abundance, singlet fermion can easily evade the stringent bounds on the spin-independent DM-nucleon direct detection cross-section due to its axial nature. However, in the allowed parameter space, strong bounds can be placed on the torsion mass and its couplings to fermions from the recent LHC searches. Assuming a non universal torsion-DM and
  torsion-SM coupling, smaller values of torsion masses may become allowed. In both cases we also study the reach of spin-dependent direct detection searches of the DM.
  
\end{abstract}



\maketitle
\flushbottom


\setcounter{footnote}{0}
\renewcommand*{\thefootnote}{\arabic{footnote}}
\section{Introduction}
\label{sec:intro}
\noindent

The Standard Model (SM) of particle physics in its present form has been extremely successful in accommodating fundamental interactions of nature. However, the model is regarded as incomplete when it comes to explain observations {\it e.g.,} dark matter (DM) abundance or the matter-anti matter asymmetry in the Universe. In ``Beyond Standard Model" (BSM) physics, one may explain 
most of these puzzles in terms of new particles and their couplings with the SM particles. But null results from  different accelerator/non-accelerator based experiments 
to test these new particles have pushed their masses above TeV scale.

The presence of DM in the universe has been established via a number of successful experiments through its gravitational impact \cite{Kolb:1990vq}
(for reviews see {\it e.g,}~\cite{Jungman:1995df,Bertone:2004pz,Feng:2010gw, Bertone:2016nfn,Lin:2019uvt}), though
its nature is yet to be known. One of the most popular and accepted DM candidate is considered to be
WIMP (Weakly Interacting Massive Particle), obtained naturally in most of the well motivated BSM scenarios like Supersymmetric (SUSY) theories or in the extra-dimensional models. However, no evidence of this prototype DM candidate has so far been observed in accelerator based experiments like the LHC or in the direct detection (DD) experiments like XENON100 \cite{Aprile:2012nq}, LUX \cite{Akerib:2016vxi}, PandaX-II \cite{Cui:2017nnn} or Xenon-1T \cite{Aprile:2018dbl} which search for DM scatterings off nuclei in detectors. All these searches have essentially put stringent constraints
on the parameter space of some of the extensions of the SM (for a review see~\cite{Arcadi:2017kky}) like
$Z$-portal~\cite{Arcadi:2014lta,Hamaguchi:2015rxa,Escudero:2016gzx,Kearney:2016rng,Balazs:2017ple}, Higgs-portal~\cite{Lebedev:2011iq,Djouadi:2011aa,Gross:2015cwa,Casas:2017jjg,Hoferichter:2017olk,Hardy:2018bph,Arcadi:2019lka}\footnote{In the context of SUSY,
lightest neutralino DM can efficiently annihilate through Higgs states to accommodate the relic  abundance~\cite{Drees:1992am,Nath:1992ty,Baer:1997ai,Ellis:2003cw,Chattopadhyay:2008hk,Chattopadhyay:2010vp,Das:2010kb}(for review see~\cite{Jungman:1995df,Bertone:2004pz,Feng:2010gw})}
$Z^\prime$-portal~\cite{Mambrini:2010dq,Dudas:2013sia,Alves:2013tqa,Lebedev:2014bba,Hooper:2014fda,Alves:2015pea,Alves:2015mua,Allanach:2015gkd,Alves:2016cqf},
or pseudoscalar portals~\cite{Berlin:2015wwa,Baek:2017vzd,Bauer:2017fsw}. Among these models, $Z^\prime$-portal with only axial coupling, or pseudoscalar portal models are somewhat preferred for a fermionic DM as non observation of any signal, specially in the context of spin-independent (SI) DM-nucleon scattering can be easily accommodated. Driven by the similar quest, here we explore the possibility of DM-SM interactions through the torsion portal which is of geometric origin, thus completely independent of any representation of the grand unified or the SM gauge group.

Due to its geometric origin, the interaction of torsion field with SM has some interesting features. In general, torsion field can be identified as one of the irreducible representations of the antisymmetric part of the
{\em affine connection} associated with the spacetime diffeomorphism symmetry. Because of its anti-symmetric property, the torsion field behaves like a massive
axial vector field when it interacts with the SM fields. In a natural extension, one may consider minimal coupling
prescription \cite{Shapiro:2001rz} which reduces the dimension of the parameter space of the theory under
consideration\footnote{As a matter of fact, superstring theories predict a non-minimal coupling of torsion with the matter fields, such that the compactification of extra-dimensions results in a general theory containing the modified version of the non-minimal couplings at low energy
  scale.\vspace{0.3cm}}. In recent times, a torsion field with masses around TeV scale has drawn attention in the
context of low energy phenomenology ~\cite{Belyaev:1997zv,Belyaev:1998ax,deBerredoPeixoto:1999vj,Das:2002qv,Mahanta:2003bf,Belyaev:2007fn,Belyaev:2016icc,Marroquim:2017cbg} (for a review see~\cite{Shapiro:2001rz}). One of this works also considers a possibility of having torsion as a DM candidate~\cite{Belyaev:2016icc} where the DMs are pair annihilated through Higgs
portal. Their analysis require vanishingly small coupling between the torsion and matter field.

In the present work, we consider a minimal model of a fermionic DM in presence of external propagating torsion field, where 
torsion plays the role of a mediator between the dark sector and the visible sector. We ignore any coupling between gauge fields and torsion fields~\cite{Shapiro:2001rz} as it may spoil gauge invariance. Since torsion appears in the covariant derivative, the coupling constants
with the DM or with the SM fields could be identical and independent of their intrinsic properties such as mass, color, flavor, charge, etc. This particular property
makes our model economical though tightly constrained from different experimental observations, particularly from the LHC searches.\footnote{However the universality in coupling constant can be broken if we consider a High scale model involving torsion. One may need to use Renormalization Group (RG) equations to calculate the coupling constants
at the electroweak scale which may be different and may provide with rich phenomenology.} Assuming a minimal deviation, i.e., allowing a non equal torsion-DM coupling compared to the SM fermions, we find that LHC
constraints can be easily accommodated. To this end we may note a few important points:

\begin{itemize}
\item Stability of the DM does {\it not} require imposition of any  {\it ad-hoc} symmetry (like ${\mathcal Z_2}$ or ${\mathcal Z_3}$ symmetry in general). Since there is no renormalizable coupling between DM-SM, the decay of DM to SM particles is automatically forbidden.

\item The presence of the axial coupling in the DM-SM interaction ensures spin-dependent (SD) direct search cross-section. In this sense, stringent limits from spin independent direct detection experiments on the DM scattering cross-section can naturally be evaded.

\item The universal coupling $\eta$ does not require any fine tuning, and can assume $\sim O(1)$ value which is
compatible with the observed DM relic abundance. In contrast, $\eta \lsim 10^{-20}$ has been assumed for torsion DM \cite{Belyaev:2016icc} scenario in order to make it stable over the cosmological time scale. This forbids any chance to produce such massive vector field at the LHC.
\end{itemize}

Our present paper is organized as follows. In Sec. II, we recapitulate the essentials of the theory of torsion and their interactions with the fermion fields. Our Sec. III  is devoted to the discussion on the experimental and theoretical constraints on the parameter space. In Sec. IV, we study the DM phenomenology of our model with torsion as a mediator between the dark and SM sector. In this analysis we particularly maintain the universality of the
torsion-fermion (this includes the fermionic DM as well) coupling parameter $\eta$, and discuss the relevant bounds on the torsion mass and its couplings to SM fermions.
We further explore the prospects of DM direct detection via a detailed scan over the relic density allowed parameter space. Direct production of the heavy resonances at the LHC may put stringent limits on the universal coupling, which can be evaded if one assumes torsion-DM coupling is different than torsion-SM fermions couplings. It is expected, as the LHC limits are mainly defined for latter interactions. A brief discussion along with changes in DM phenomenology would be presented in Sec. V. Finally, we summarize our key results in Sec. VI.

\section{ Origin of Torsion and its coupling with matter}
\label{sec:prelim}

\noindent 
We begin this section with a brief introduction to torsion. Here we closely follow Refs.\cite{Shapiro:2001rz,Belyaev:1997zv,Belyaev:1998ax,deBerredoPeixoto:1999vj,Belyaev:2007fn,Belyaev:2016icc}. In the general theory of relativity, the covariant derivative of a vector field $A^\nu$ is a tensor and is defined as:
\bea
\nabla_\mu \, A^\nu =     \partial_\mu \, A^\nu + \Gamma^\nu_{\,\,\mu\lambda}\,A^\lambda,                    
\eea
where the last term $\Gamma^\lambda_{\,\,\mu\nu}$ is known as the {\it affine connection}. The simplest form of this affine connection in general relativity is known as the Christoffel symbol and is based on the two following special requirements
\begin{itemize}
\item it is symmetric in nature, $\Gamma^\lambda_{\,\,\mu\nu} = \Gamma^\lambda_{\,\,\nu\mu}$,
\item  the metricity condition  i.e., $\nabla_\alpha \, g_{\mu\nu} = 0$,
\end{itemize}
where $g_{\mu\nu}$ is the metric tensor of the corresponding spacetime we are working in. When the above two conditions are fulfilled we have an expression of the Christoffel symbol as,
\bea
\Gamma^\lambda_{\,\,\mu\nu} = \frac{1}{2}\, g^{\lambda\alpha}\, \big(\partial_\mu\, g_{\nu\alpha} +  \partial_\nu\, g_{\mu\alpha}  - \partial_\alpha\, g_{\mu\nu}\big).
\eea  
However, in the Einstein-Cartan extension of general relativity, the said conditions can be relaxed so that the affine connection can have an anti-symmetric part defined as:
\bea
{\tilde\Gamma}^\lambda_{\,\,\mu\nu} = {\Gamma}^\lambda_{\,\,\mu\nu} + C^\lambda_{\,\,\mu\nu},
\eea

where $C^\lambda_{\,\,\mu\nu}$ is a mixed tensor field. Thus, new degrees of freedom are introduced into the system which cannot be fixed by the background metric anymore. Therefore, even in the flat space limit, this field will survive. A particular irreducible representation of this tensor field may play the role for a torsion field.

For better understanding, we may define the general torsion tensor as
\bea
T^\lambda_{\,\,\mu\nu}~=~{\tilde\Gamma}^\lambda_{\,\,\mu\nu} - {\tilde\Gamma}^\lambda_{\,\,\nu\mu},
\eea
which is antisymmetric in its lower indices. The above torsion tensor can be re-written in terms of irreducible representation: 
\bea
T_{\mu\nu\lambda} = \frac{1}{3} \, \big( T_{\nu}\,g_{\mu\lambda} - T_{\lambda}\,g_{\mu\nu} \big) -
\frac{1}{6} \,\varepsilon_{\mu\nu\lambda\sigma}\,S^{\sigma} + q_{\mu\nu\lambda},
\eea
where $S^\lambda = \varepsilon^{\mu\nu\alpha\lambda}\,T_{\mu\nu\alpha}$ is torsion trace axial vector mode, $q_{\mu\nu\alpha}$ is trace free tensor mode satisfying the conditions $q^{\mu}_{\,\,\nu\mu} = 0$ and $\varepsilon^{\mu\nu\alpha\sigma} \, q_{\mu\nu\alpha} = 0$. The trace free vector mode is defined as $T_\mu =  T^{\alpha}_{\,\,\mu\alpha}$. Because of its geometric nature and specific parity properties, different irreducible torsion components will have specific type of couplings with the fermions. The general action of a Dirac
spinor coupled to torsion field reads:
\bea
{\cal S}_D \, = \, \int d^4 x \, \left\{\,i \bar{f} \gamma^\mu \big(\partial_\mu - i e \, A_\mu
- i \eta_1 \, \gamma^5 \, S_\mu  + i \eta_2 \,T_\mu \big)f - m \,\bar f \, f \right \} ,
\eea
where $\eta_1$, $\eta_2$ are non-minimal coupling parameters. Simply from the symmetry properties of various terms of the torsion fields, in the minimal coupling scenario, these parameters may assume fixed values: $\eta_1=\frac{1}{8}$ and $\eta_2=0$. However, if we consider  dynamical torsion interactions with the matter fields, $\eta_1$ and $\eta_2$ are expected to change under the renormalization group (RG) flow (see\ e.g.,\cite{Shapiro:2001rz} and references therein). For phenomenological purposes we choose them as free parameters at low energy scale. Moreover, keeping the simplicity in mind, we will set all the other irreducible components of the torsion field to be zero except the traceless axial vector component $S_{\mu}$. 

We consider massive fermion fields in the action mentioned above. In the fermion mass limit $m=0$, the action enjoys an additional gauge symmetry transformation related to torsion:
\bea
f \to f^\prime = f\,e^{\gamma_5\beta(x)}
,\,\,\,\,\,\,\,\,\,\,\,\,\,\,
{\bar {f}} \to {\bar {f}}^\prime 
= {\bar {f}}\,e^{\gamma_5\beta(x)}
,\,\,\,\,\,\,\,\,\,\,\,\,\,\,
S_\mu \to S_\mu^\prime = S_\mu - {\eta}^{-1}\, \partial_\mu\beta(x)\,.
\label{chiral}
\eea
Massive fermion terms may be explained if the said
symmetry is softly broken. This in turn will automatically generate the torsion mass term at the loop level, and will be proportional the fermionic mass term. This fact prompts one to consider a massive dynamical torsion field $S_\mu$
with the following simplest action (see \cite{Shapiro:2001rz}, for details)
\bea
{\cal S}_{torsion} = \,\int d^4 x \,\Big\{\,-\frac14\,S_{\mu\nu}S^{\mu\nu} + \frac12\,m_s^2\, S_\mu S^\mu\,\Big\}\,,
\eea
where $S_{\mu\nu} = \partial_\mu\,S_\nu - \partial_\nu\, S_\mu$ and $m_s$ is the torsion mass. The first term is the
usual abelian kinetic term which naturally satisfies the above gauge
conditions in Eq.\ref{chiral}.

Inclusion of a DM, specifically a SM-singlet Dirac fermion $\psi$, is possible in a similar way since, by construction, the new state interacts with torsion field in the same way like that of a SM fermion. Importantly, the singlet Dirac state cannot have any renormalizable couplings to any SM particles. Thus DM-SM interaction can be only realized through $S_\mu$ portal, which reads:
\bea
    {\cal S}_{DM} \, = \, \int d^4 x \, \left\{\,i \bar{\psi} \gamma^\mu
      \big(\partial_\mu - i \eta \,
      \gamma^5 \, S_\mu \big)\psi - m_{\psi} \,\bar\psi \, \psi \right \}.
    \eea
\noindent
Finally, the relevant parts of the action which includes the new interactions can be written as:

    \bea
    {\cal S}_{New} &=& {\cal S}_{torsion} + {\cal S}_D + {\cal S}_{DM} = \,\int d^4 x \,\Big\{\,-\frac14\,S_{\mu\nu}S^{\mu\nu} + \frac12\,m^2_s\,
    S_\mu S^\mu\,\Big\} \nonumber\\ 
&+& \int d^4 x\, \big[\left\{\,i \bar{f} \gamma^\mu \big(\partial_\mu - i \eta_1 \, \gamma^5 \, S_\mu \big)f \right \} +
      \left\{\,i \bar{\psi} \gamma^\mu \big(\partial_\mu - i \eta \, \gamma^5 \, S_\mu \big)\psi- m_{\psi} \,\bar\psi \,
      \psi \right \}\big].
\label{action_comp}
\eea

In our analysis we will mainly consider two scenarios:

\begin{itemize}

\item Universal coupling scenario ($\eta=\eta_1$): Both SM and DM can couple to the torsion field with a universal coupling $\eta$, which is a natural outcome of torsion portal. The model, thus naturally becomes very predictive as only a minimal set of input parameters are required to compute the DM phenomenology.
\item
  Non-universal coupling (NU) scenario ($\eta \ne \eta_1$) : Though the universal coupling scenario is natural, but can be highly constrained in terms of the
  available parameter space. We shall illustrate this in detail in the
  subsequent sections. The situation may be considerably improved if one relaxes the universality condition, and set only the DM-torsion coupling $\eta$ as free parameter. We shall also briefly discuss the outcome of such a scenario where $\eta$ is different from torsion-SM coupling $\eta_1$.
\end{itemize}

\section{Constraints on the model parameters}
\label{sec:const}

In this section we will discuss the constraints arising from various experimental and theoretical bounds. Under the universal coupling scenario (this is always be the case unless otherwise stated), our model can be characterized by only three free parameters, namely:

\bea
\{m_s, m_{\psi},\eta\},
\label{eq:freeparam}
\eea

which respectively are the torsion mass, the DM mass and the torsion-matter universal coupling. The constraints on the DM mass come from the requirement of obtaining
observed relic abundance and satisfying the direct detection bounds. We discuss these in detail in the next section. Here we will mainly summarize the experimental constraints on the torsion mass and its coupling focusing on the recent LHC results.

\begin{itemize}
\item Before the advent of LHC, torsion mass and couplings were not entirely free parameters, rather significantly constrained from different analysis, including accelerator searches. Assuming the torsion mass to be considerably heavier than other particles present in the action (Eq.~\ref{action_comp}), torsion interaction with fermion fields can be effectively recasted in terms of four-fermion contact interaction~\cite{Belyaev:1997zv,Belyaev:1998ax}.
\bea
\mathcal L_{int}=-\frac{\eta^2}{m_s^2}(\bar{f} \gamma^\mu \gamma^5 f)(\bar{f} \gamma^\mu \gamma^5 f).
\eea 

Now, such contact interactions can be constrained from several experiments as elaborated in~\cite{Belyaev:1997zv,Belyaev:1998ax}. From global analysis one can put a limit on the contact axial $eeqq$ interaction, which then translates into a bound on the torsion mass and its coupling~\cite{Belyaev:1997zv,Shapiro:2001rz}:
\bea
\frac{m_s}{\eta}> 1.7~{\text {TeV}}.
\label{eq:tormass}
\eea
For a light torsion with mass below 1 TeV, accelerator searches from LEP or TEVATRON collaborations became important. Analysis of the data constraints the coupling $\eta$ in the range $0.02\lsim\eta\lsim 0.1$~\cite{Belyaev:1997zv,Belyaev:1998ax}, depending upon torsion mass. A combined analysis from LEP-2 and TEVATRON data excludes torsion mass upto 600 GeV unless the torsion-fermion coupling drops below 0.1. Also, for torsion coupling ($\sim 0.3-0.6$), the torsion mass is pushed well above TeV scale.  

In recent years both ATLAS and CMS collaborations
have searched for high-mass resonances through their direct production and subsequent decays to dilepton, dijet or mono-$X$ final states~\cite{Aad:2012hf,Sirunyan:2016iap,Sirunyan:2017hci,Aaboud:2017phn,Aaboud:2017yvp,Sirunyan:2018exx,Goldouzian:2627472,White:2019eqh,Aad:2019fac,Sirunyan:2019vgt}. Dilepton
final states are of special interests as they provide the most stringent bounds, though other searches namely dijet and mono-jet limits \cite{Sirunyan:2016iap,Aaboud:2017yvp,Aaboud:2017phn,Sirunyan:2017hci} may also be of importance~\cite{Drees:2019iex}. Based on the
dilepton searches, that includes the recent CMS analysis at 13 TeV center of mass (CM) energy with integrated luminosity 140 fb$^{-1}$~\cite{CMS-PAS-EXO-19-019}, a bound on $m_s \gsim 4.5$~TeV can be placed assuming fermion coupling to torsion is different than the respective couplings with the $Z$ boson. In fact, a similar limit for
$\eta \sim 0.1$ was earlier presented in~\cite{Belyaev:2007fn} where authors analyzed LHC discovery reach in the $\eta-m_s$ plane at 14 TeV CM energy with integrated luminosity of 100 fb$^{-1}$. However, they considered only SM fermions in their analysis. Torsion specific analysis has also been performed by the ATLAS collaboration at $\sqrt s = 7$~TeV, for the same $e^-$ and $\mu^-$ final states and the bound is much relaxed compared to the present one. One may also refer to the latest ATLAS $t\bar t$ production results to set limits on torsion parameters considering torsion-top quark couplings $O(1)$~\cite{Marroquim:2017cbg}. Here a much weaker constraint $m_s \gsim 3$~TeV can be observed.

In this analysis, our primary objective is to study the interplay of different input parameters in DM phenomenology considering universal coupling scenario in particular. So, on the very outset, we do not impose any collider constraint on the parameter space. After investigating
the DM phenomenology, limits are placed to depict the net available parameter space. In applying the LHC bounds, one should keep in mind that such limits on the torsion mass have been derived assuming narrow width approximation (NWA) while studying the distribution of the invariant mass of the decay products. However, this approximation does not hold good in the limit $\Gamma_s/m_s\geq 0.5$~\cite{Abdallah:2014hon,Alves:2015pea}, where perturbative limits break down. In this regime direct application of LHC limits may not be completely justified. Keeping this in mind we indicate the part of the parameter space where indeed one may have $\Gamma_s/m_s\geq 0.5$.  

\item The propagating axial vector field $S_\mu$ may also contribute to the muon anomalous magnetic moment ($a_\mu$) at one loop, 
though the contribution is found to be negative:
$a^S_\mu (\rm {m_s>>m_\mu})\sim -\frac{\eta^2}{4\pi^2}\frac{5 m_\mu^2}{3 m_s^2}$ ~\cite{Das:2002qv,Queiroz:2014zfa}. From the recent measurements of $a_\mu$~\cite{PhysRevD.98.030001} one may observe the discrepancy in $\Delta a_\mu$ is positive: $\left(268\pm 76.27\right)\times 10^{-11}$ which is an excess of $\sim 3.51\sigma$ compared to the SM expectation. Thus, only a positive contribution from the BSM physics can match the difference which is not favoured in the present case. 

\item A theoretical constraint can arise from the requirement of the preservation of unitarity of the theory due to quantum corrections~\cite{Belyaev:1998ax,Belyaev:1997zv,deBerredoPeixoto:1999vj,Belyaev:2016icc}. This demands:
\bea
\frac{m_s^2}{\eta}>>m_f^2,
\label{eq:qcorrect}
\eea
where $m_f$ is the mass of any fermion field that couples to the torsion. This constraint, however, is satisfied for our entire parameter space. 

\end{itemize}
\section{Dark Matter through Torsion portal: WIMP scenario}
\label{sec:dmpheno}

As explained earlier, we consider the DM phenomenology of a  singlet Dirac fermion $\psi$ in presence of a torsion field where interactions between the SM fermions or the DM with the torsion is parameterized by a universal coupling parameter $\eta$. We remind that the DM is stable over the time scale of the age of the universe without requiring any ad-hoc symmetry. This is also true in the parameter space where DM is heavier than the torsion field.

\begin{figure}[htb!]
$$
\includegraphics[scale=0.4]{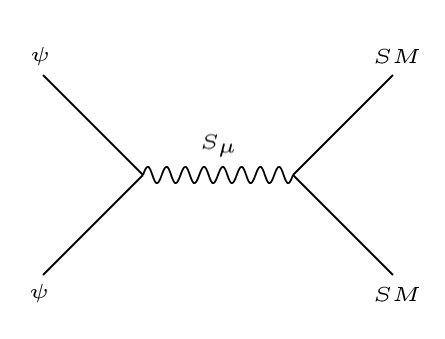}
\includegraphics[scale=0.45]{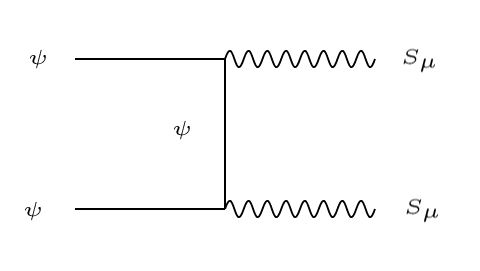}
$$
\caption{Pair-annihilation of the DM $\psi$ to the SM particles
  via torsion-portal (left) and to the torsion pairs via $t$-channel process (right).}
\label{fig:annihil}
\end{figure}

In the WIMP paradigm, the DM particles are assumed to be in thermal equilibrium with the SM in the early universe. In the present set-up, the thermal equilibrium is maintained through the torsion mediated $s$-channel and $t$-channel processes as shown in Fig:\ref{fig:annihil}. At some later time when the rate of interaction becomes smaller than the Hubble rate of expansion the DM falls out of the equilibrium, and this defines freeze out~\cite{Kolb:1990vq,Jungman:1995df} of the DM:

\bea
\Gamma_{A}< H,
\eea

where $\Gamma_{A}$ is the interaction rate of the DM pair annihilations and it is given by:

\bea
\Gamma_{A} = n_\psi \langle\sigma v\rangle, 
\eea

where $n_\psi$ is the DM number density and $H$ is the Hubble constant. The present number density of the DM is decided by its number density at the time of freeze-out, which can be computed by using the standard prescription of Boltzmann equation~\cite{Gondolo:1990dk} for a single component DM candidate. Note that, the relic abundance of the DM is solely controlled by the $2\to 2$ annihilation channels, where the annihilation cross-sections are dependent on the parameters in Eq.~\ref{eq:freeparam}. In the present case, dominant non-relativistic form of the thermally-averaged annihilation cross-section for the $s$ channel
($\psi \bar \psi \to \bar f f $) and $t$ channel ($\psi \bar \psi \to S_\mu S_\mu$ when kinematically allowed)
processes can be expressed as:

\bea
\langle\sigma v\rangle_{\bar ff}\simeq \sum_f N_c^{f} \frac{\eta^4}{2\pi}\frac{m_f^2}{m_s^4}
\sqrt{1-\frac{m_f^2}{m_\psi^2}}\left(\frac{\left(m_s^2-4 m_\psi^2\right)^2}
     {\left(m_s^2-4 m_\psi^2\right)^2+\Gamma_s^2 m_s^2}\right) + \mathcal{O}\left(v^2\right),
\label{schannel}
\eea

\bea
\langle\sigma v\rangle_{ss}\simeq \frac{\eta^4}{16\pi m_\psi^2} \left(1-\frac{m_s^2}{m_\psi^2}\right)^{3/2}
\left(1-\frac{m_s^2}{2 m_\psi^2}\right)^{-2} + \mathcal{O}\left(v^2\right),
\label{tchannel}
\eea

where $v$ is the relative velocity of the annihilating DM pair, $m_{f}$ corresponds to the mass of all SM fermions that couple to the torsion and $N^f_c$ refers the number of colors for final state fermions (= 3(1) for quarks (leptons)). $\Gamma_s$ is the total decay width of the torsion to all fermions given by (in the rest frame of the torsion):

\bea
\Gamma_s=\frac{N_c^f\eta^2m_s}{12 \pi}
\left[1-4(\frac{m_f}{m_s})^2\right]^{\frac{3}{2}} + \eta^2 \frac{m_s}{12 \pi}
\left[1-4(\frac{m_\psi}{m_s})^2\right]^{\frac{3}{2}},
\label{gammatot}
\eea
where the last term is relevant for $m_s > 2m_\psi$.
The cross-section has been computed for $s=4m_\psi^2$
($s$ is the CM energy). 
$\mathcal{O}\left(v^2\right)$ refers to the velocity dependent $p$-wave terms, which are sub-dominant in presence of $s$-wave. It is intriguing to note that for $s$-channel process, assuming $m_\psi \sim m_s >> m_f$, the dominant $v^2$ contribution involves $\frac{\eta^4 m_\psi^2}{\left(4m_\psi^2-m_s^2\right)^2}$, which carries the resonance effect~\cite{Berlin:2014tja,Arcadi:2017hfi}. Interestingly the leading order term in Eq.\ref{schannel} may not be sensitive to Breit-Wigner type narrow width resonance. The $p$-wave terms can also lead to important contributions for the relic density calculations, specially when the DM is lighter than top quark. Then the contribution of the leading order {\it i.e,} the velocity independent term in the $s$ channel (see Eq.\ref{schannel}) is extremely suppressed $\sim \frac{m_f^2}{m_s^4}$, where $f$ refers any fermions except the top quark. The $p$-wave term becomes significant as it goes as $\sim \frac{1}{m^2_\psi}$ for $m_\psi \sim m_s$. Similarly $t$-channel contributions, free from $v^2$ suppression, are also important for $m_s \lsim m_\psi$. For numerical calculations, we use the general expressions for $\langle\sigma v\rangle$. This has been done by implementing the model in {\tt CalcHEP}~\cite{Belyaev:2012qa} and then the model output has been fed to public code {\tt MicrOmegas}~\cite{Belanger:2006is,Belanger:2008sj,Belanger:2013oya,Barducci:2016pcb}. The valid parameter space should comply with the observed relic abundance data~\cite{Jarosik:2010iu,Aghanim:2018eyx}:

\bea
\Omega_{\psi}^{obs}h^2=0.1199\pm 0.0022.
\label{relic}
\eea

\begin{figure}[htb!]
$$
\includegraphics[scale=0.4]{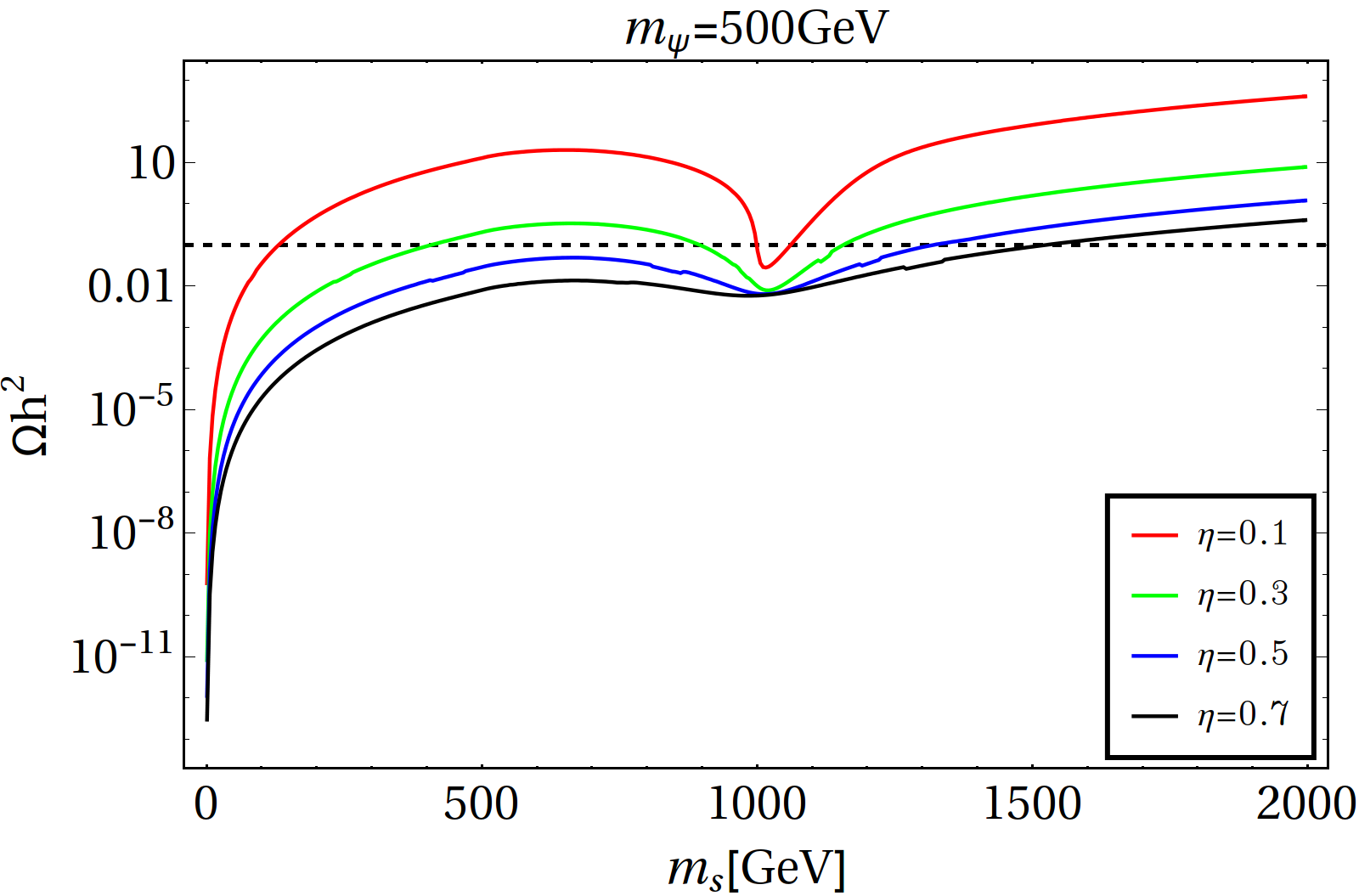}
\includegraphics[scale=0.4]{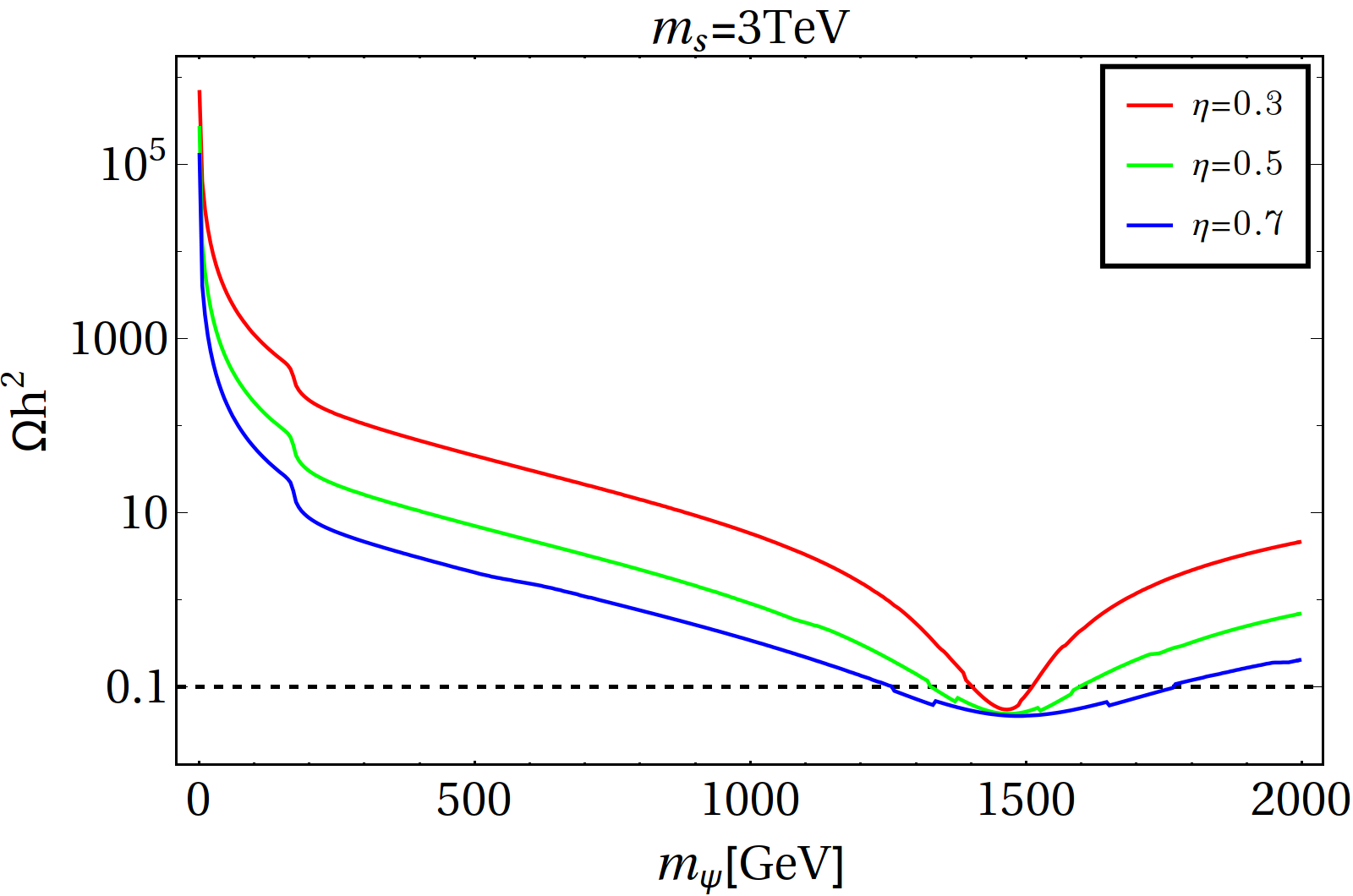}
$$
\caption{Left: Variation of relic abundance of $\psi$ with $m_s$ for different choices of the couplings $\eta:\{0.1,0.3,0.5,0.7\}$, are shown in red, green, blue and black respectively. Here the DM mass is set at 500 GeV. Right: Same for a fixed torsion mass $m_s=3~\rm TeV$ for
three different choices of $\eta:\{0.3,0.5,0.7\}$ respectively in red, green and blue. In each figure the black dashed line corresponds to the PLANCK observed relic abundance.}
\label{fig:relic1}
\end{figure}

For illustration we first fixed the DM mass $m_\psi=500$~GeV and plotted the variation of relic abundance with $m_s$ for four different choices of the $\eta:\{0.1,0.3,0.5,0.7\}$ respectively which
are shown by red, green, blue and black curves on the left hand side (LHS) of Fig.~\ref{fig:relic1}. As expected, in each case we see a resonance at $m_\psi \simeq \frac{m_s}{2}$ where the observed relic abundance~\cite{Aghanim:2018eyx} is satisfied. The relic abundance increases for smaller $\eta$, and at resonance it is difficult to produce an acceptable $\Omega h^2$ for $\eta< 0.1$.  The resonance regions may become flat for a larger $\eta$ when the Breit-Wigner decay width $\Gamma_s$ is large for a given torsion mass $m_s$. The situation is different for a lighter $S_\mu$ hence for a lighter DM, as both $s$ and $t$ channel processes may be relevant here. As an example, for our chosen DM mass, $\langle\sigma v\rangle_{\bar\psi\psi\to \bar t t}$ can be sufficiently large even for a moderate value of $\eta$, where right relic abundance or even a under abundant region can be observed. Here annihilations are facilitated to  all possible final states. 
The viable parameter space may also include the region where $m_s \lsim m_{\psi}$. Next, we fix the torsion mass $m_s=3~\rm TeV$  and choose three different couplings $\eta:\{0.3,0.5,0.7\}$ as shown on the right hand side (RHS) of Fig.~\ref{fig:relic1} through red, green and blue curves respectively. Here the relic density is satisfied only at the resonance region while larger width can be observed for a larger $\eta$.

\begin{figure}[htb!]
$$
\includegraphics[scale=0.4]{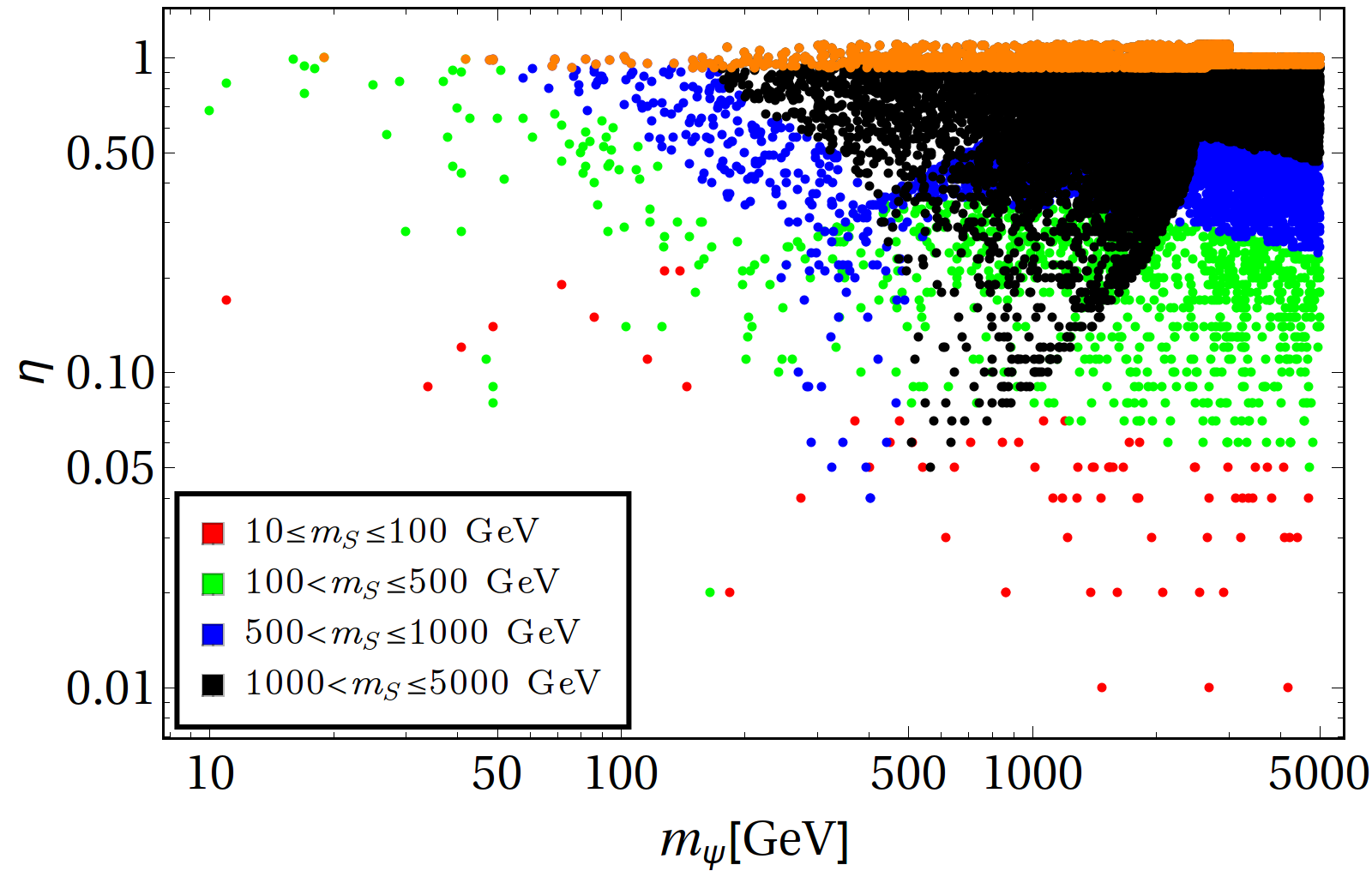}
\includegraphics[scale=0.4]{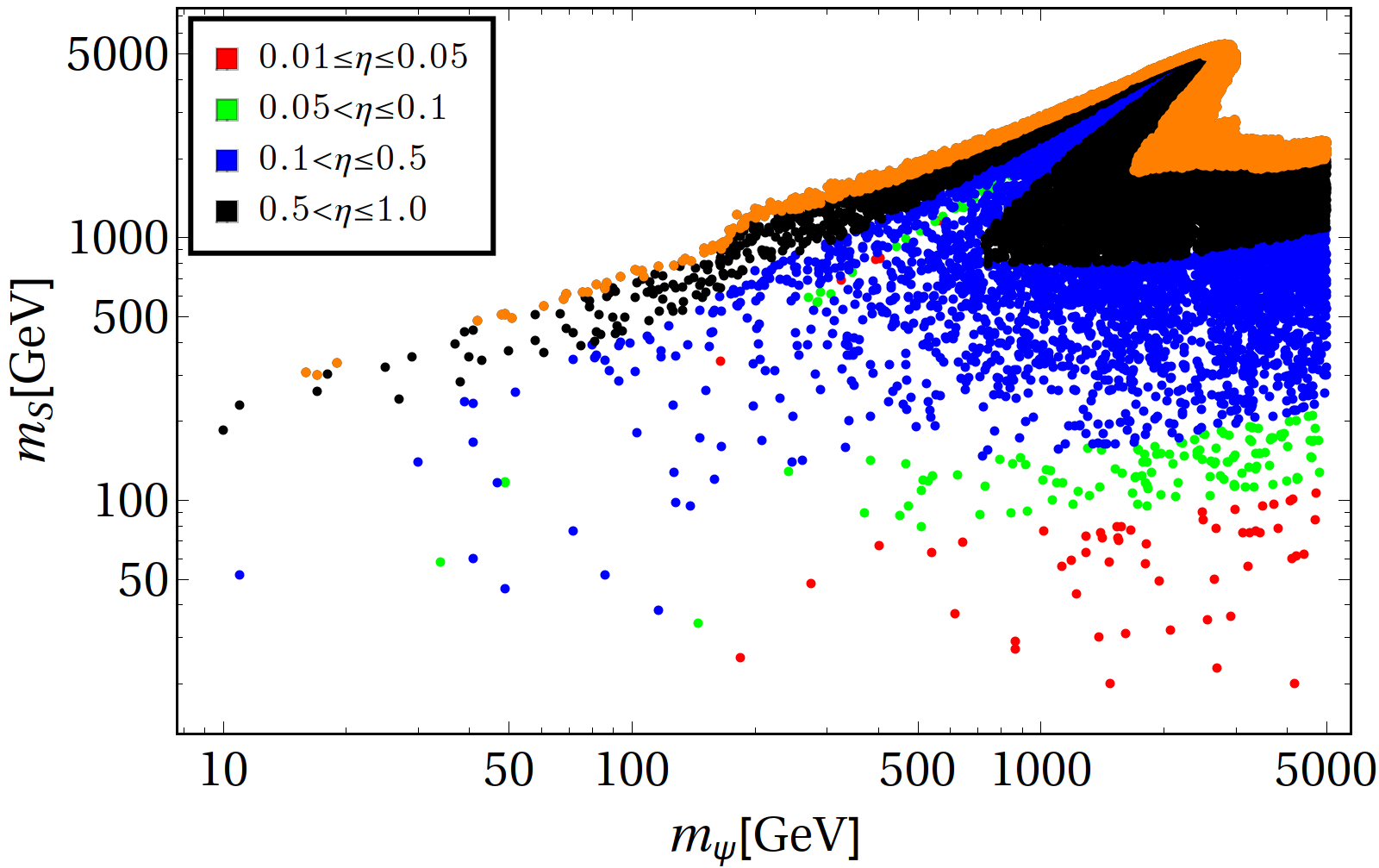}
$$
\caption{Left: Relic density satisfied parameter space in $m_{\psi}$-$\eta$ plane, where different colours show torsion masses in the range $\{10-100\}~\rm GeV$ in red, $\{100-500\}~\rm GeV$ in green, $\{500-1000\}~\rm GeV$ in blue and $\{1000-5000\}~\rm GeV$ in black. Right: Parameter space satisfying relic abundance in $m_\psi$-$m_s$ plane,
where the colour codes corresponding to different values of the coupling   $\eta$: \{0.01-0.05\} in red, \{0.05-0.1\} in green, \{0.1-0.5\} in blue and \{0.5-1.0\} in black. In each plot the orange region corresponds to $\Gamma_s/m_s\geq 0.5$ limit.}
\label{fig:relscan}
\end{figure}

Now, in order to explore the parameter space satisfying the observed relic abundance (via Eq.\ref{relic}) we scan over the parameters listed in Eq.\ref{eq:freeparam} in the
following range:

\bea
m_{\psi}:\{10-5000~{\text{GeV}}\};m_s:\{10-10^4~{\text{GeV}}\}; \eta:\{0.01-1.0\}.
\eea

The corresponding parameter space is shown in Fig.~\ref{fig:relscan}. On the LHS we show the parameter space satisfying relic abundance in $m_\psi$-$\eta$ plane for different ranges of the torsion mass $m_s$ indicated by different colours. Here both DM and the torsion can take small mass values $\sim$10 GeV in different parts
of the parameter space specified by $\eta$, where relic abundance can be satisfied either through annihilations to $f\bar f$ or via $t$ channel process. 

In the RHS of Fig.~\ref{fig:relscan}, relic density allowed region is shown in $m_\psi-m_s$ plane where different colours correspond to different values of the coupling $\eta$. We see here, smaller values of $\eta$ (in red) favours $m_s\lsim 100~\rm GeV$, while a larger $m_s$ is required for larger $\eta$ (in blue and black) to satisfy the relic abundance. This is expected as the rise in the torsion mass reduces $\langle\sigma v\rangle$. Hence larger $\eta$ is expected to make relic density compatible with the observed abundance. In each plot highlighted orange regions depict $\Gamma_s/m_s\geq 0.5$, where LHC constraints may not be directly applied to disqualify a parameter space point as discussed in Sec.~\ref{sec:const}.

\subsection{Indirect detection of the Dark Matter}
\label{sec:indir}

The relic density allowed parameter space for the DM can also be probed via different indirect detection experiments that look for annihilation and/or decay products of the DM that produces SM particles. Residual DM annihiliations may still
occur at late universe where the relative velocity takes a value $\sim 10^{-3}$.
Among these final states, photon and neutrinos, being neutral and stable, can reach the detectors without getting much affected by the intermediate space. In case of high energy $\gamma$-rays, one considers hadronization and production of neutral pions and its subsequent decays to two photons with varying energies and also final state radiations.  
Detectable gamma-rays can be tested at the instruments like the
Fermi
Large Area Telescope (LAT)\footnote{http://fermi.gsfc.nasa.gov}
in space or 
the ground-based Cherenkov telescopes like
MAGIC \footnote{https://magic.mpp.mpg.de/}.
Since no signal has so far been observed, we would like to see the status of
the relic abundance satisfying parameter space from the bounds arising from a
joint global analysis from the MAGIC and the Fermi-LAT which search for
gamma-ray signals from DM annihilation in dwarf spheroidal galaxies (dSphs)~\cite{Ahnen:2016qkx}.

\begin{figure}[htb!]
$$
\includegraphics[scale=0.4]{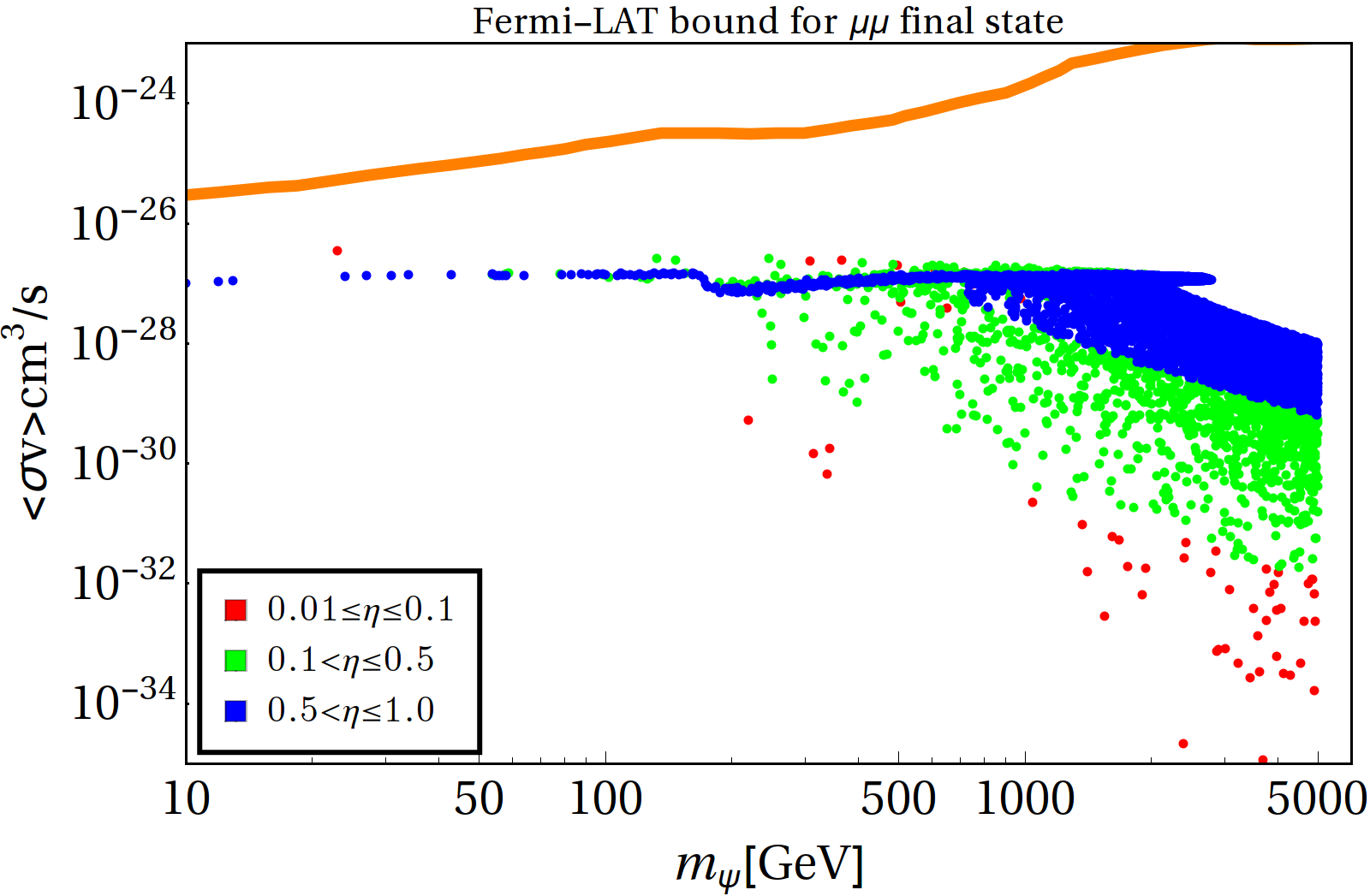}
\includegraphics[scale=0.4]{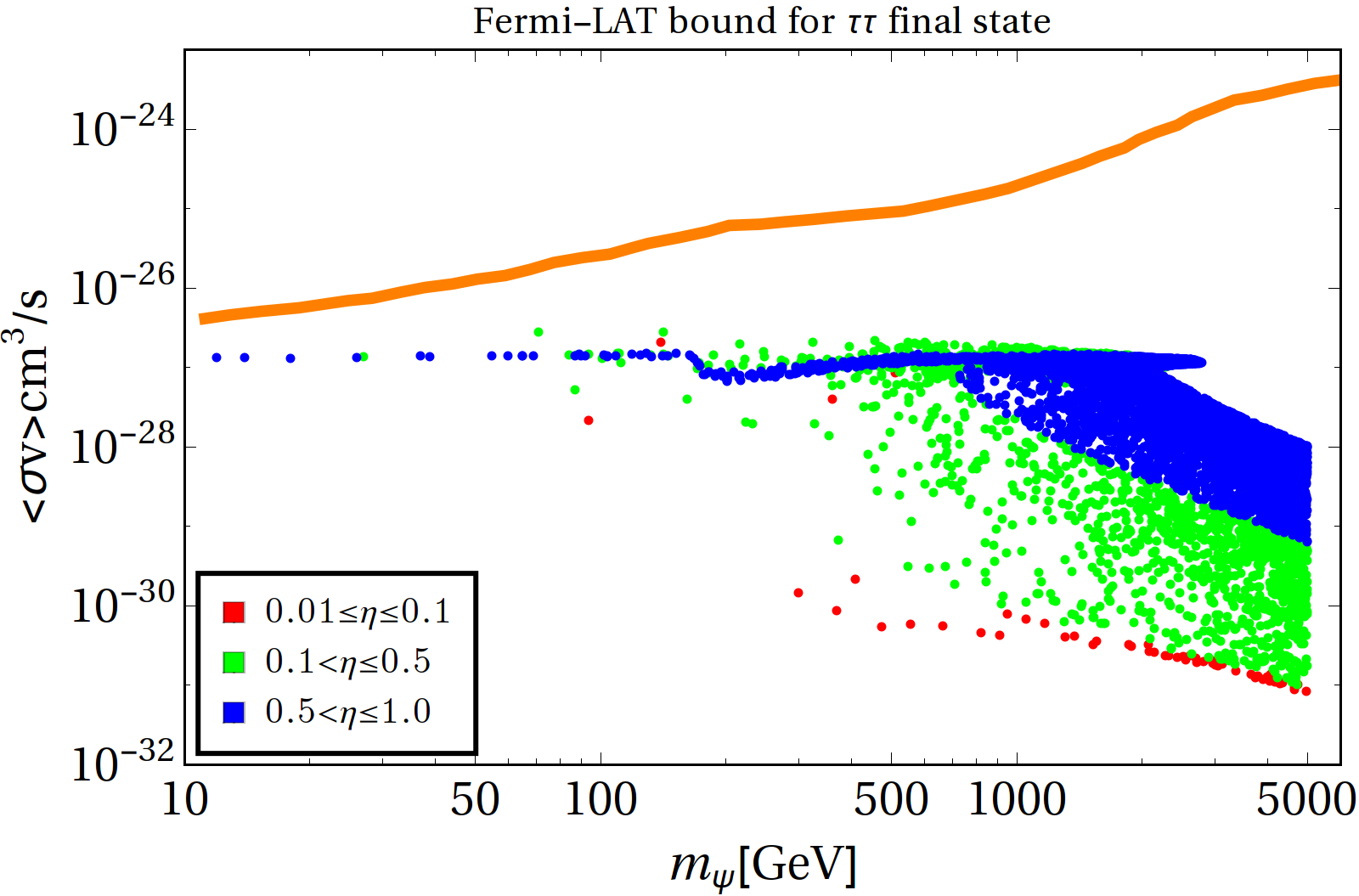}
$$
$$
\includegraphics[scale=0.4]{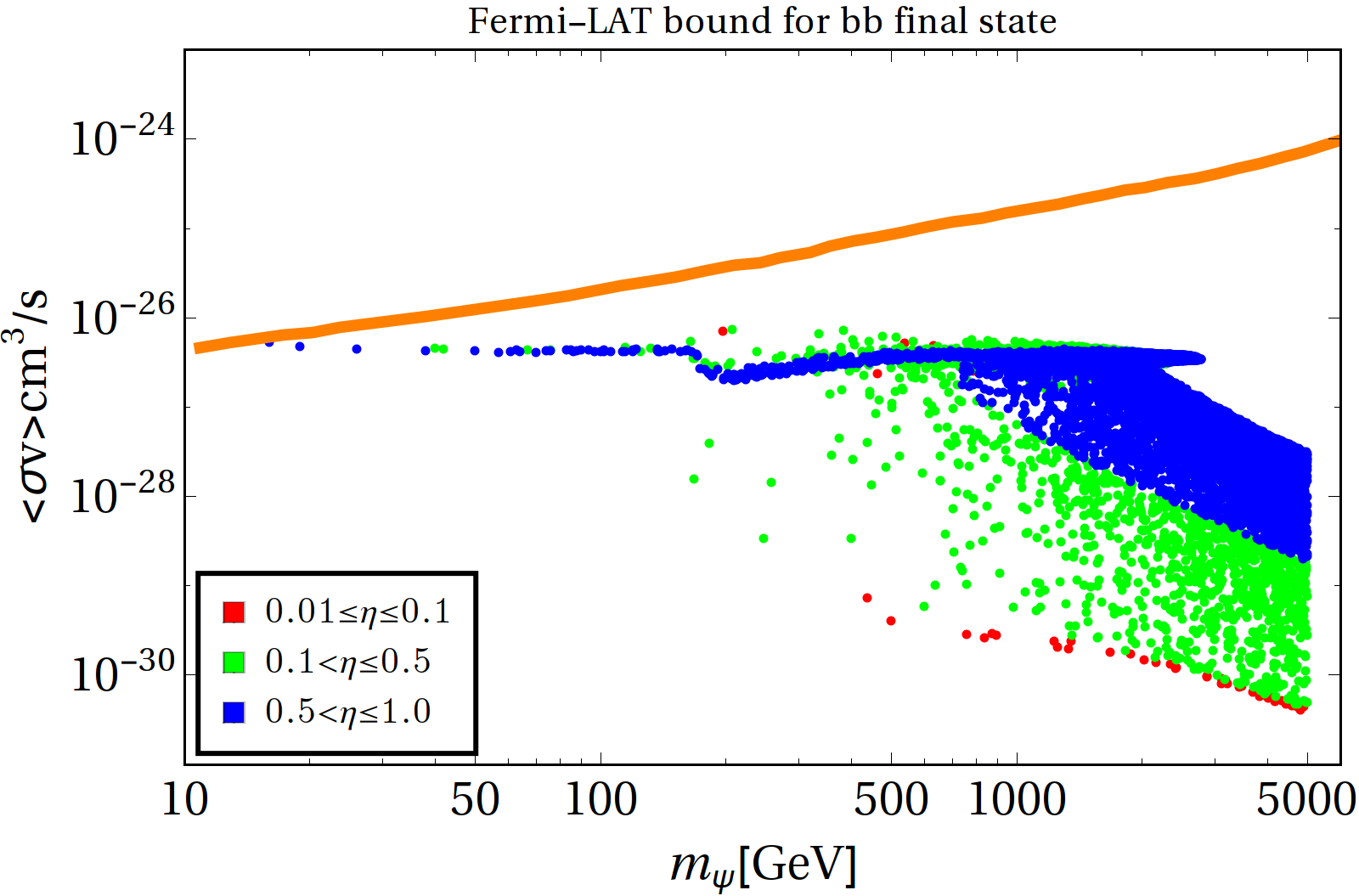}
$$
\caption{DM annihilations into $\mu^+\mu^-$ (top left), $\tau^+\tau^-$
  (top right) and $b\bar b$ (bottom) compared against indirect detection
  bounds from the combined observation of MAGIC and Fermi-LAT. In each
  figure regions corresponding different couplings are shown in
  red \{0.01-0.1\}, green \{0.1-0.5\},
  blue \{0.5-1.0\}. The thick orange curve represents bound from
  non observations of $\gamma$-rays corresponding to particular final state.}
\label{fig:indir}
\end{figure}

The  differential gamma ray flux produced due to WIMP-like DM annihilation in a given region of the sky $\left(\Delta\Omega\right)$ and observed at the Earth
reads~\cite{Cirelli:2010xx,Ibarra:2013cra,Gaskins:2016cha,Slatyer:2017sev,Hooper:2018kfv}:

\bea
\frac{d\Phi}{dE}\left(\Delta\Omega\right) = \frac{1}{4\pi} \frac{\langle\sigma v\rangle J\left(\Delta\Omega\right)}{2 m_\psi^2}\frac{dN}{dE},
\eea

where $dN/dE$ is the average gamma-ray spectrum per annihilation and,

\bea
J\left(\Delta\Omega\right) = \int_{\Delta\Omega}d\Omega^{'} \int_{LOS} dl \rho^2\left(l,\Omega^{'}\right), 
\eea

is the well-known $J$-factor with $\rho$ being the DM density profile, and the integrals running over $\Delta\Omega$ and the line-of-sight (LOS) through
the DM distribution. Here we consider a Navarro-Frenk-While (NFW) profile for the computation \cite{Navarro:2008kc}.
Since the DM $\psi$ can annihilate into charged fermion final states: $\mu^+\mu^-,\tau^+\tau^-,b\bar b$ etc, all such final states can give rise to gamma rays. Thus, one can constrain the DM annihilations to different charged final states from non observations of gamma rays. In Fig.~\ref{fig:indir} we bounded our relic density allowed parameter space from MAGIC and Fermi-LAT data ~\cite{Ahnen:2016qkx} for muon (top left), tauon (top right) and $b$-quark (bottom) final states. We see that for our choice of the coupling $\eta$ and torsion mass $m_s$, the parameter space is completely unconstrained. This is not completely unexpected,
as the velocity dependent contributions in Eq. \ref{schannel} which may show
the resonance behaviour is largely suppressed at late times.


\subsection{Direct Detection of the Dark Matter}
\label{sec:dd}

\begin{figure}[htb!]
\includegraphics[scale=0.35]{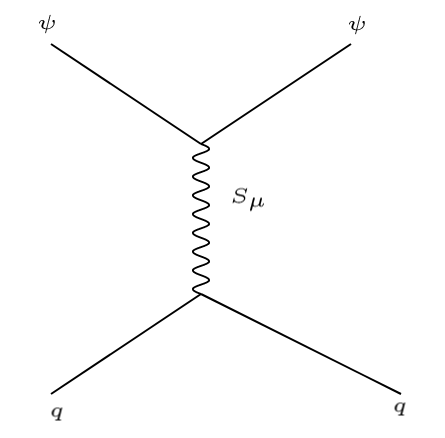} 
\caption{DM-nucleon elastic scattering mediated by the torsion.}
\label{fig:dd}
\end{figure}

Since the DM can annihilate into the SM particles via $s$-channel mediation of the torsion producing right relic abundance, the same interaction can give rise to DM-nucleon elastic scattering. This will lead to the direct detection of the DM in present and future experiments. In the present case, driven by a purely axial coupling, the DM-nucleon scattering would give rise to only spin-dependent
interaction. Note that, the SD cross section involves the spin content of the nucleus i.e., there is a $J(J+1)$ enhancement from the nuclear spin $J$. This is highly welcome since a weak-scale Dirac dark matter accompanied by
vector like interactions with the some mediator $M_\mu$ {\it i.e,} $(\psi \gamma^{\mu} \psi) M_{\mu}$ is either ruled out or very tightly constrained.
The relevant Feynman graph for the spin-dependent
scattering is depicted in Fig.~\ref{fig:dd}. For such a process, the spin-dependent direct search cross-section is proportional to ~\cite{Belanger:2008sj}:

\bea
\sigma_{SD} \sim \frac{\eta^4}{m_s^4}\frac{16}{\pi}\mu^2\left(\frac{J+1}{J}\right)\left(\xi_p\langle s_p\rangle+\xi_n\langle s_n\rangle\right)^2
\label{eq:ddxsec},
\eea

\begin{figure}[htb!]
$$
\includegraphics[scale=0.4]{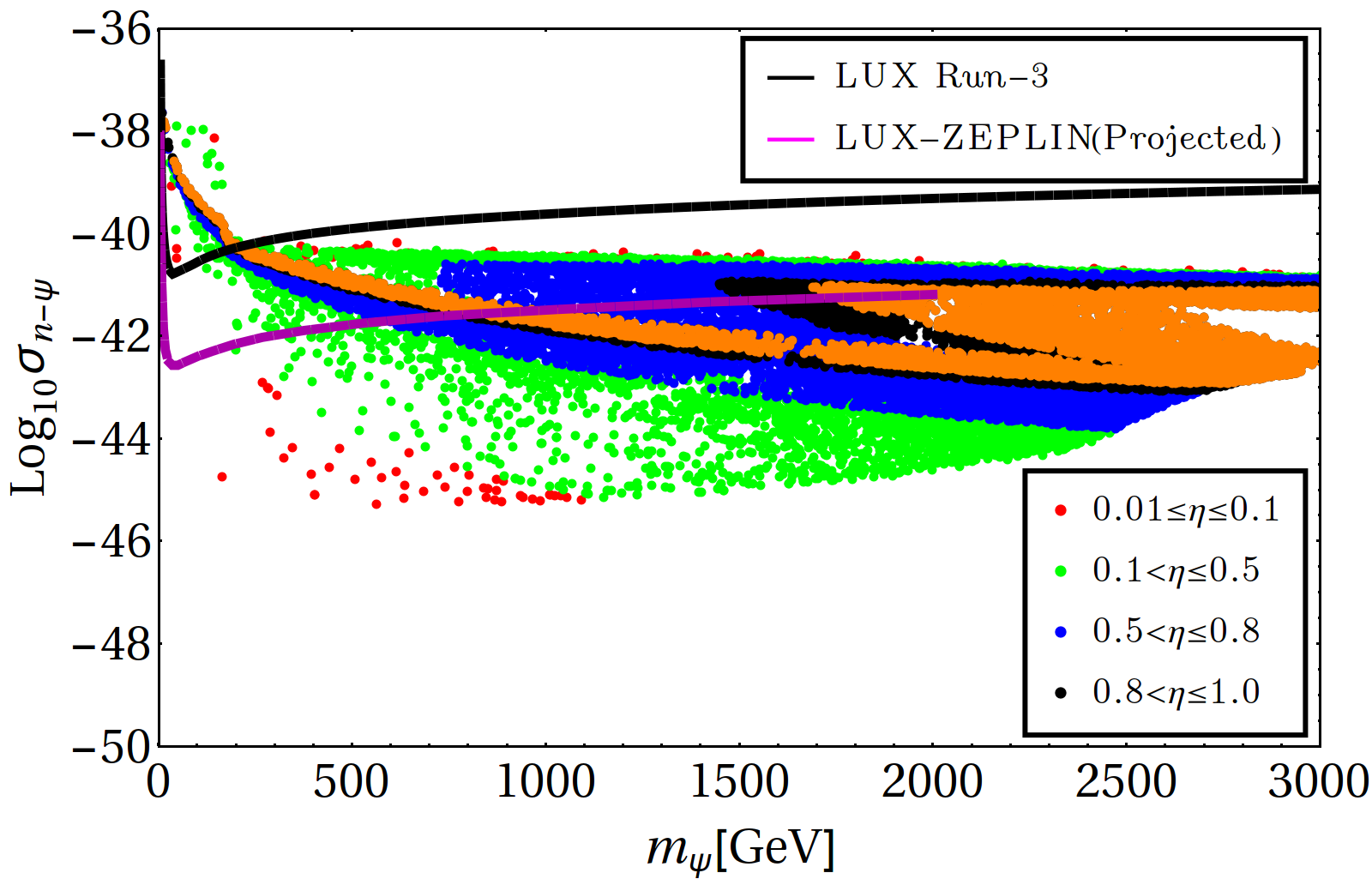}
\includegraphics[scale=0.4]{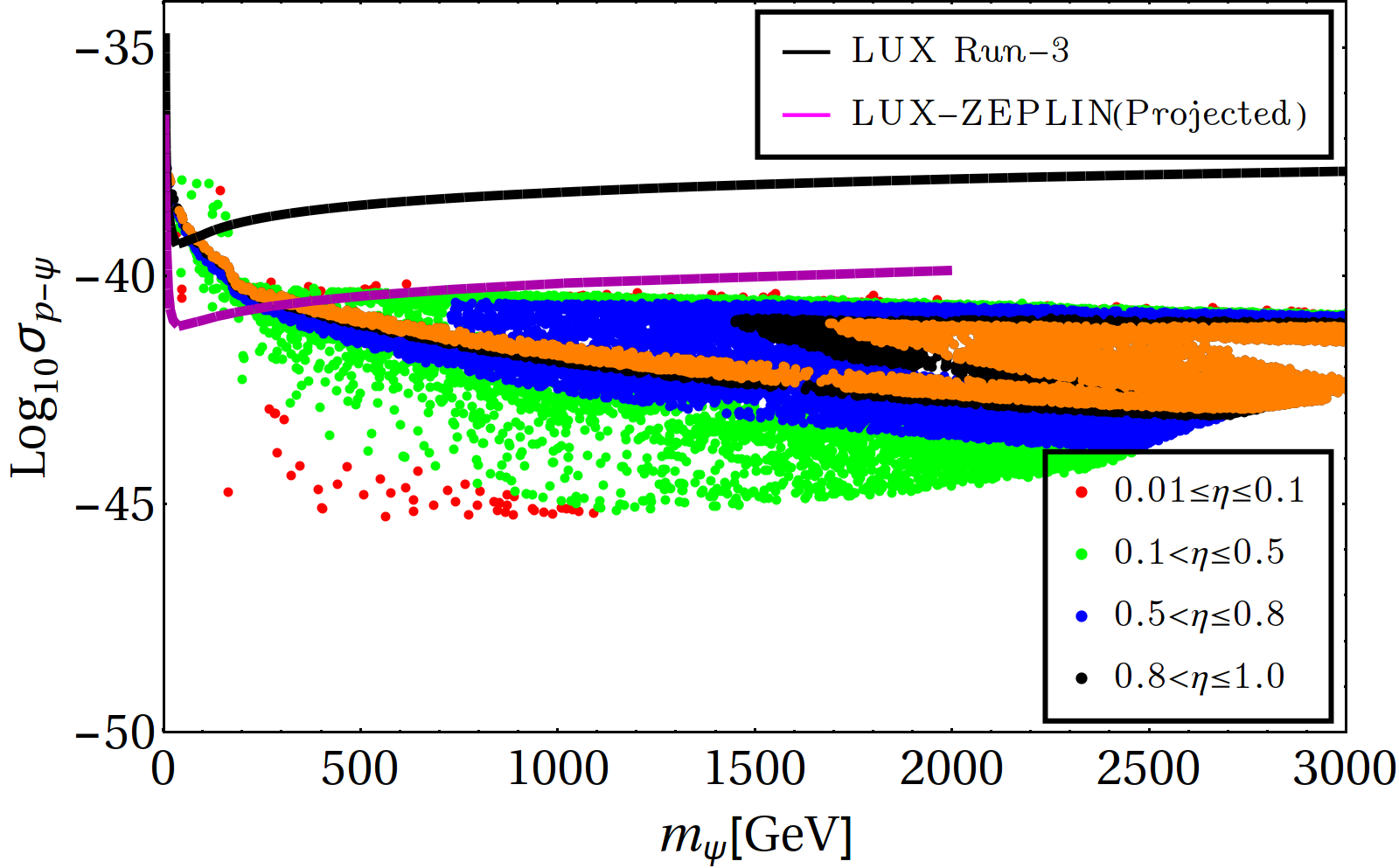}
$$
$$
\includegraphics[scale=0.32]{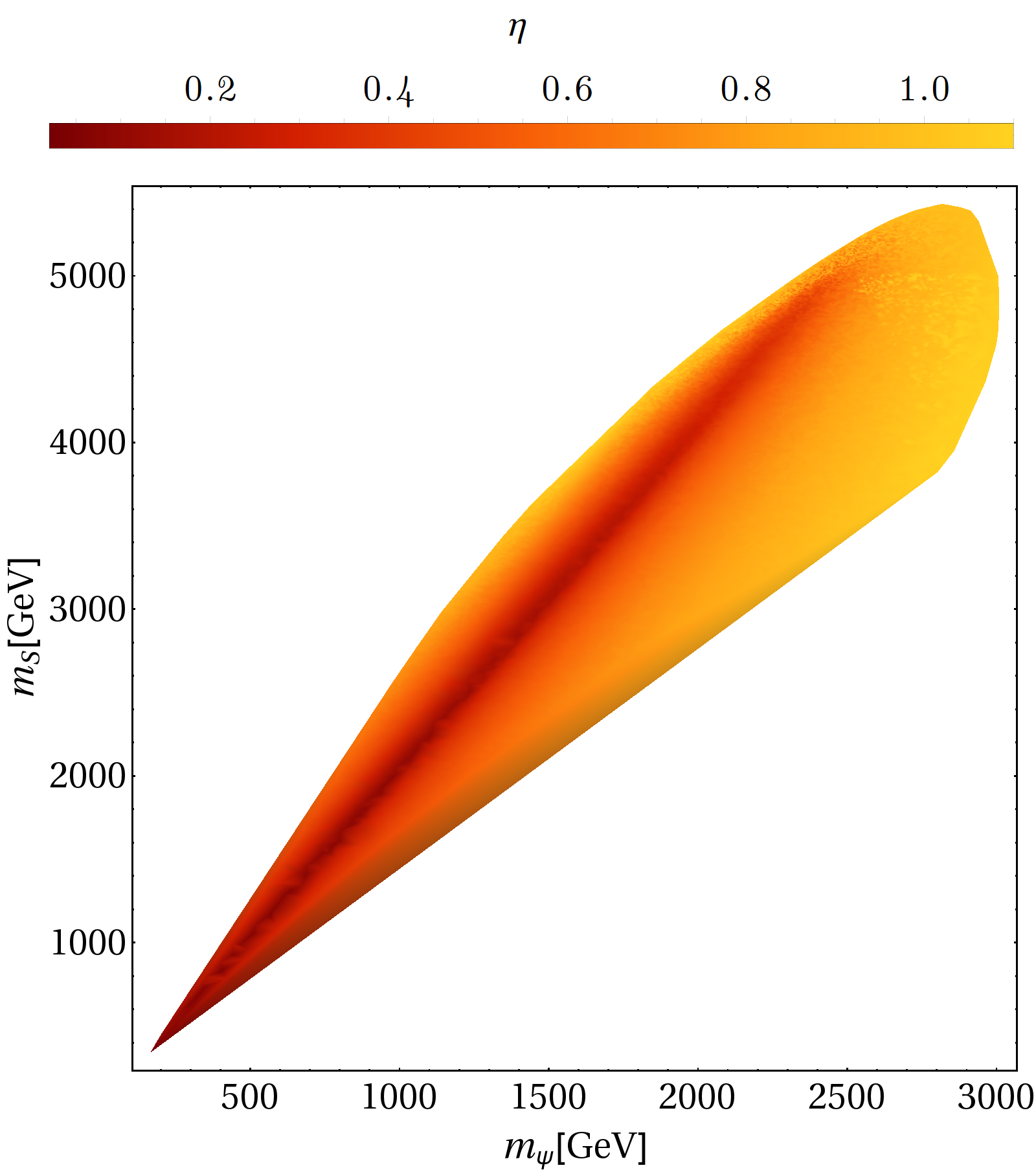}
\includegraphics[scale=0.32]{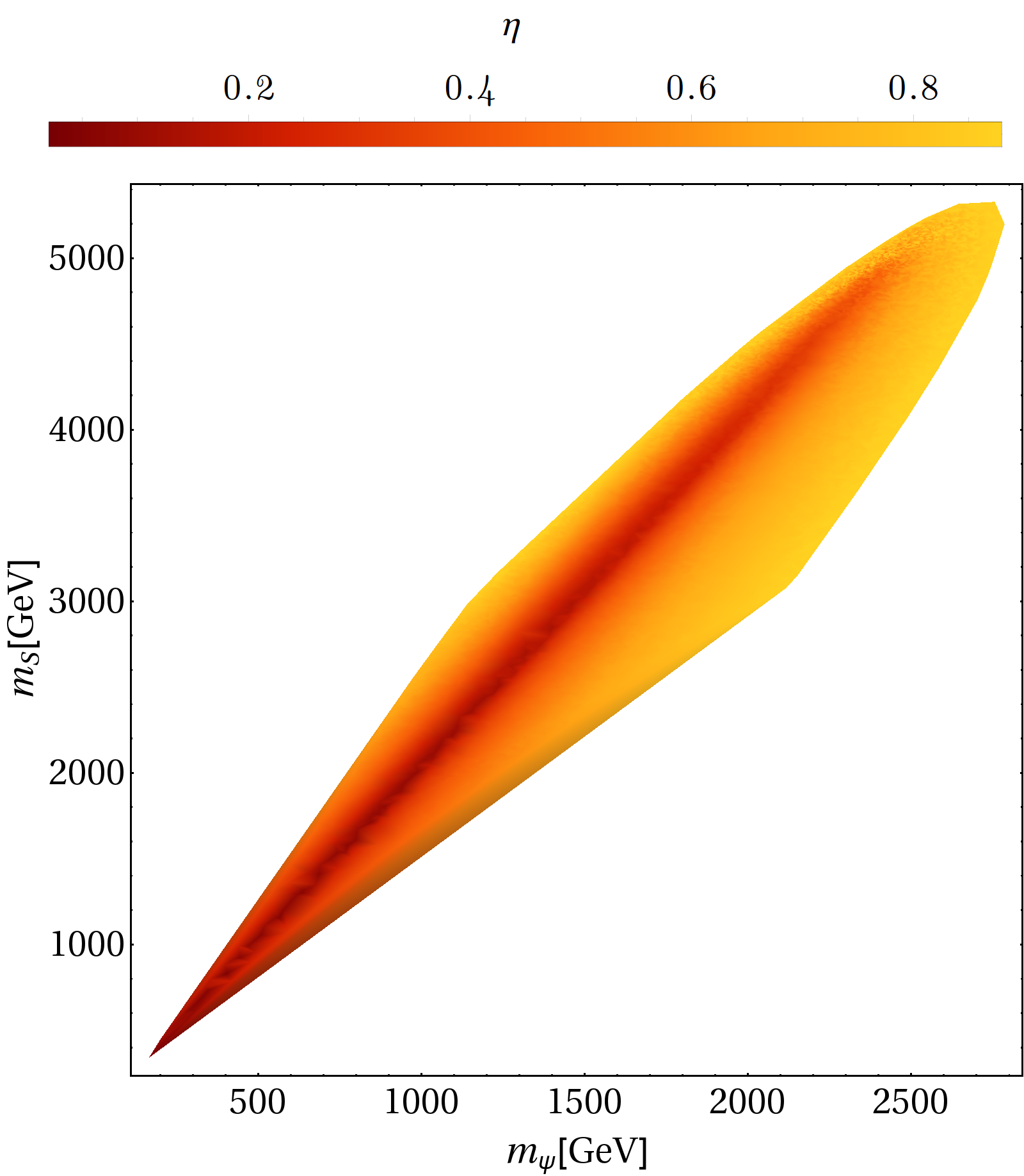}
$$
\caption{Top Left: Relic density satisfied parameter space in direct search
  plane, where different colours correspond to different choices of
  $\eta$: \{0.01-0.1\} in red, \{0.1-0.5\} in green, \{0.5-0.8\} in blue and
  \{0.8-1.0\} in black. The black dashed line shows the exclusion limit from
  spin-dependent direct search for neutron-DM scattering. Similarly, the
  magenta line refers future LZ sensitivity. Top Right: Same
  but here the exclusion limits refer spin-dependent
  direct search for proton-DM scattering. In both plots orange
  region corresponds to $\Gamma_s/m_s\geq 0.5$. Bottom Left: Net parameter
  space satisfying relic abundance and spin-dependent direct search in
  $m_\psi$-$m_s$ plane, where the top panel shows the variation of the coupling
  $\eta$. Bottom Right: Net parameter space within the perturbative limit
  $\Gamma_s/m_s\lsim 0.5$.}
\label{fig:ddrel}
\end{figure}
\vspace{1cm}

where $\mu=\frac{m_\psi m_N}{m_\psi+m_N}$ is the DM-nucleon reduced mass with $N\in\{ n,p\}$ {\it i.e,} neutron or proton. $\langle s_{p,n}\rangle$ are obtained from nuclear calculations or from simple nuclear models, such as the odd-group model~\cite{Belanger:2008sj} and they are estimated to be $\approx 0.5$ for a nuclei with an odd number of protons or neutrons, while zero for nuclei with even number of protons or neutrons. $\xi_{p,n}$ are obtained from lattice data~\cite{Airapetian:2006vy,Ageev:2007du,Belanger:2008sj}. We consider the most recent limit on spin-dependent WIMP-proton or WIMP-neutron cross-section from the run-3 data of the LUX experiment~\cite{Akerib:2016lao} and constrain our relic density satisfying parameter space. In both the plots DM and torsion mass span over a few TeV mass scale and $\eta:\{0.01-1.0\}$.
From the top panel in Fig.~\ref{fig:ddrel} we see that the relic density allowed parameter space lies well below the present bound from LUX. This is due to the additional
suppression coming from the torsion mass in Eq.~\ref{eq:ddxsec}. Similarly, smaller coupling results in even smaller cross-section.
In the plots, we also show projected sensitivity of the next generation LUX-ZEPLIN (LZ) direct detection experiments which may in future probe some parts of the parameter space \cite{Akerib:2018dfk}. Clearly, SD neutron-DM scattering offers a better prospect which can be attributed to
even number of protons present in the xenon-based detectors like LUX \cite{Schnee:2011ef}. Finally in the bottom left panel of Fig.~\ref{fig:ddrel} we show the net
parameter space satisfying relic abundance and direct search exclusion limit. We find that if we strictly follow $\Gamma_s/m_s\lsim 0.5$, the parameter space is shrinked a bit in the high torsion mass region with $\eta \lsim 0.8$.

\vskip 1.0cm
\section{Limits on the universal coupling scenario and non universal torsion
   matter coupling}
\label{nonuni}

\begin{figure}[htb!]
$$
\includegraphics[scale=0.5]{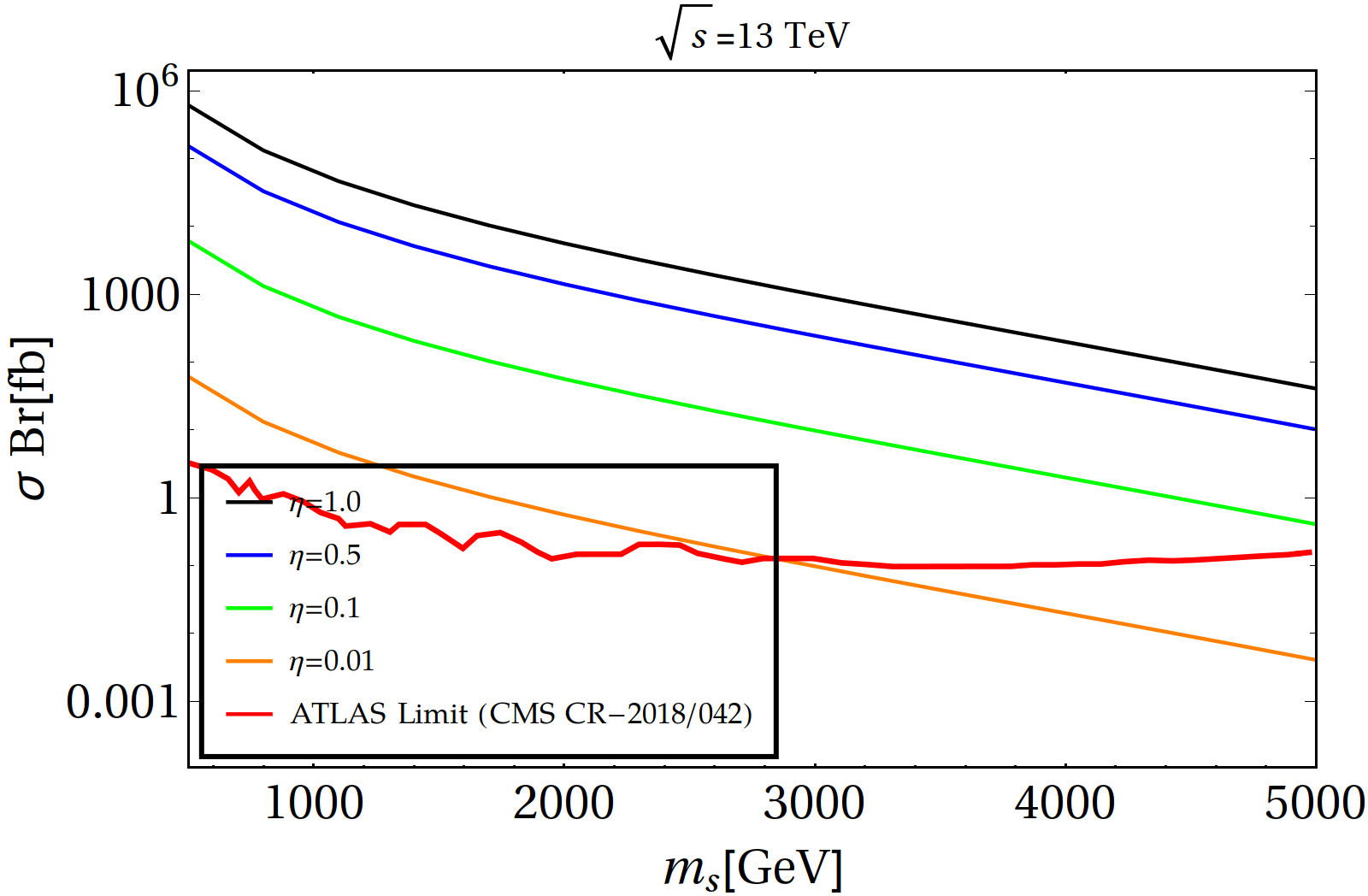}
$$
\caption{$\sigma (pp \to S_\mu) \times Br(S_\mu \to l^+ l^-)$ ($l \in e,\mu$) vs. $m_s$ in the universal coupling scenario. }
\label{fig:nu}
\end{figure}

Based on our discussion in Sec.~\ref{sec:const}, it is quite expected that our minimal coupling scenario would be tightly bounded from the direct production of torsion and its subsequent decays to charged leptons: $e,\mu$. For a qualitative understanding, we calculated effective cross-section for dilepton productions $\sigma (pp \to S_\mu) \times Br(S_\mu \to l^+ l^-)$ ($l \in e,\mu$) and plotted the variation with $m_s$ for different sets of torsion-matter couplings. Torsion production cross-section $\sigma (pp \to S_\mu)$ at the LHC has been calculated using {\tt MadGraph-v2.6.6}~\cite{Alwall:2011uj} at a CM energy of 13TeV with PDF choice {\tt CTEQ6L} \cite{Alwall:2011uj,Alwall:2014hca,Placakyte:2011az,Ball:2012cx,Ball:2014uwa} while the branching fraction has been determined from Eq.~\ref{gammatot}. The result is presented in Fig.~\ref{fig:nu}, where the exclusion limit is from~\cite{Goldouzian:2627472}. The bound would be relaxed if one considers finite width of the mediator \cite{White:2019eqh}. Torsion mass in the range $m_s \sim \{3-5\}$~TeV is typically allowed modulo the universal coupling $\eta$ is small $\eta:\{0.01-0.1\}$. However, in this range, satisfying the PLANCK data will be extremely difficult as one requires relatively light $S_\mu$ to compensate the coupling suppression in order to produce relic abundance in the right ballpark. In other words, the typical $m_s$ value which can satisfy the observed abundance for a given $\eta$ in the aforesaid range has been mostly ruled out by the LHC results. On the higher side of the coupling $\eta \sim 1$, one mostly finds $\frac{\Gamma}{m_s} > 0.5$, thus LHC results cannot be used to rule out any parameter space point.

\begin{figure}[htb!]
$$
\includegraphics[scale=0.4]{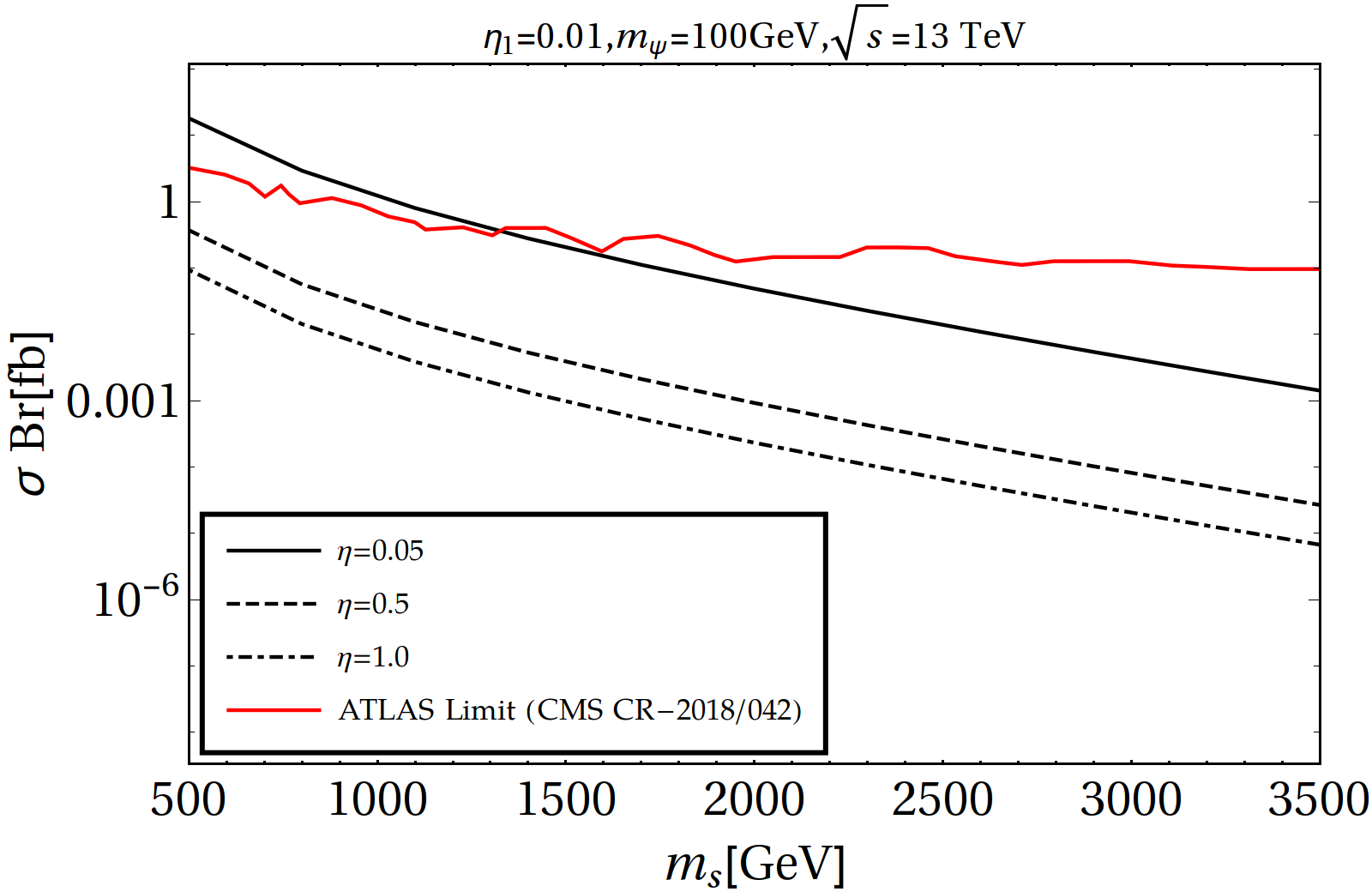}
\includegraphics[scale=0.4]{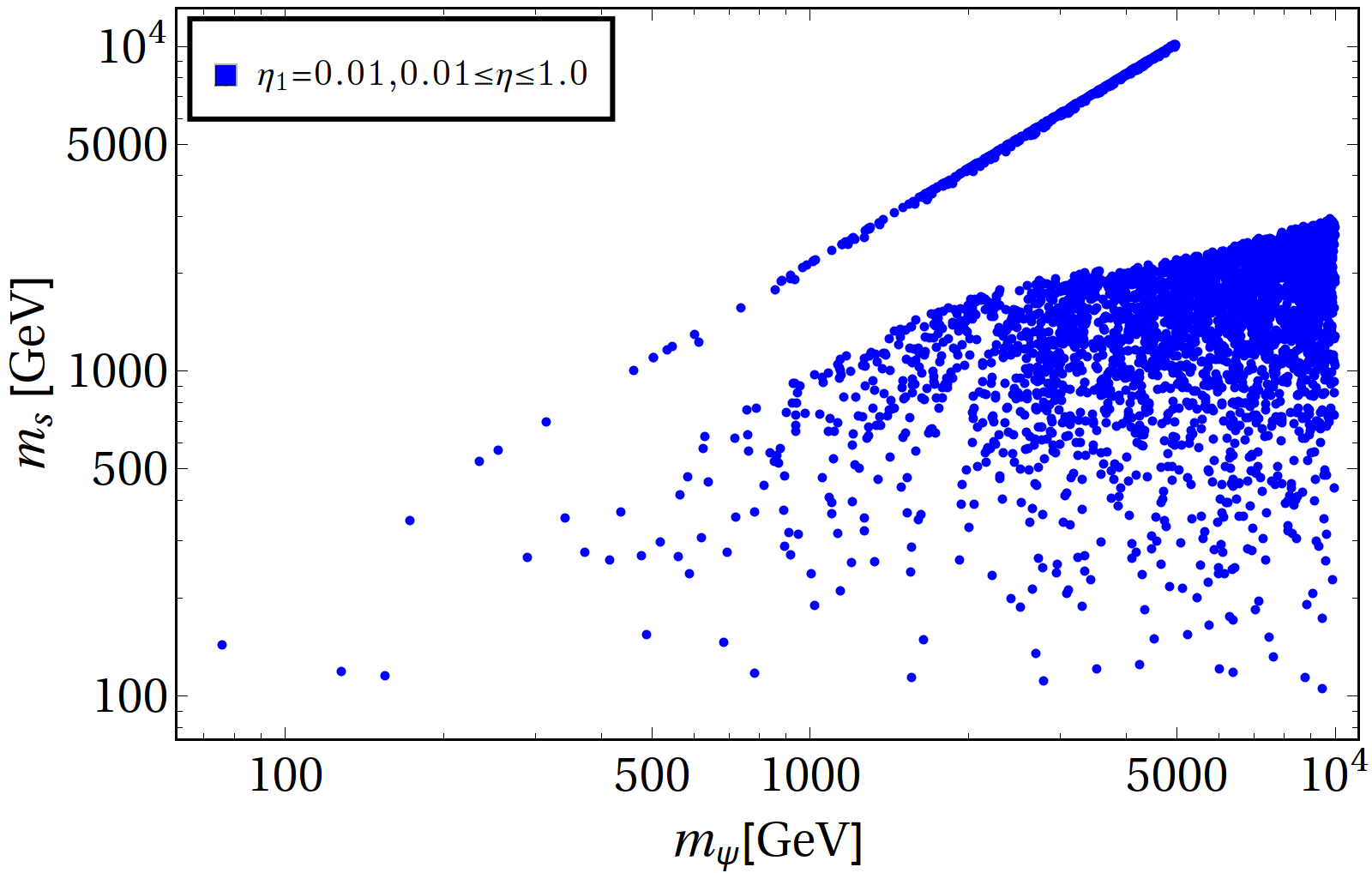}
$$
\caption{Left: $\sigma (pp \to S_\mu) \times Br(S_\mu \to l^+ l^-)$
  ($l \in e,\mu$) vs. $m_s$ in the non minimal coupling scenario. Here
  torsion-SM fermion couplings are set at $\eta_1=0.01$. In the minimal
  coupling case, the bound on $m_s$ reads as $m_s > 3$~TeV. But with a
  different torsion-DM coupling, torsion's branching fraction to leptonic final states will be reduced further to allow even lighter $S_\mu$. Assuming $\eta=0.05$, one finds $m_s > 1.4$~TeV using LHC dilepton searches. Right:
  Allowed parameter space complied with PLANCK data in $m_s-m_\psi$ plane in the NU coupling scenario.}
\label{fig:nulimits}
\end{figure}

\begin{figure}[htb!]
$$
\includegraphics[scale=0.4]{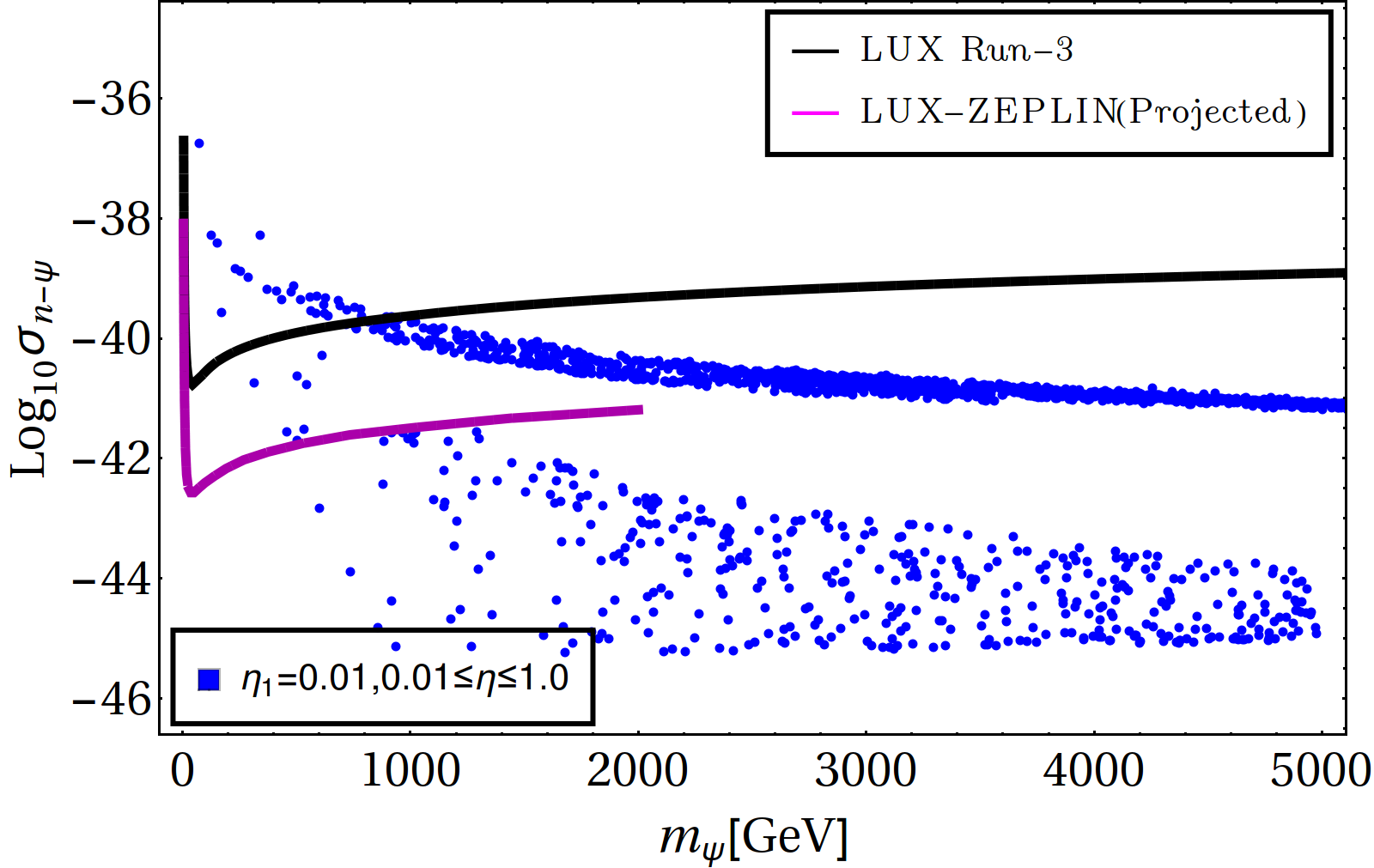}
\includegraphics[scale=0.4]{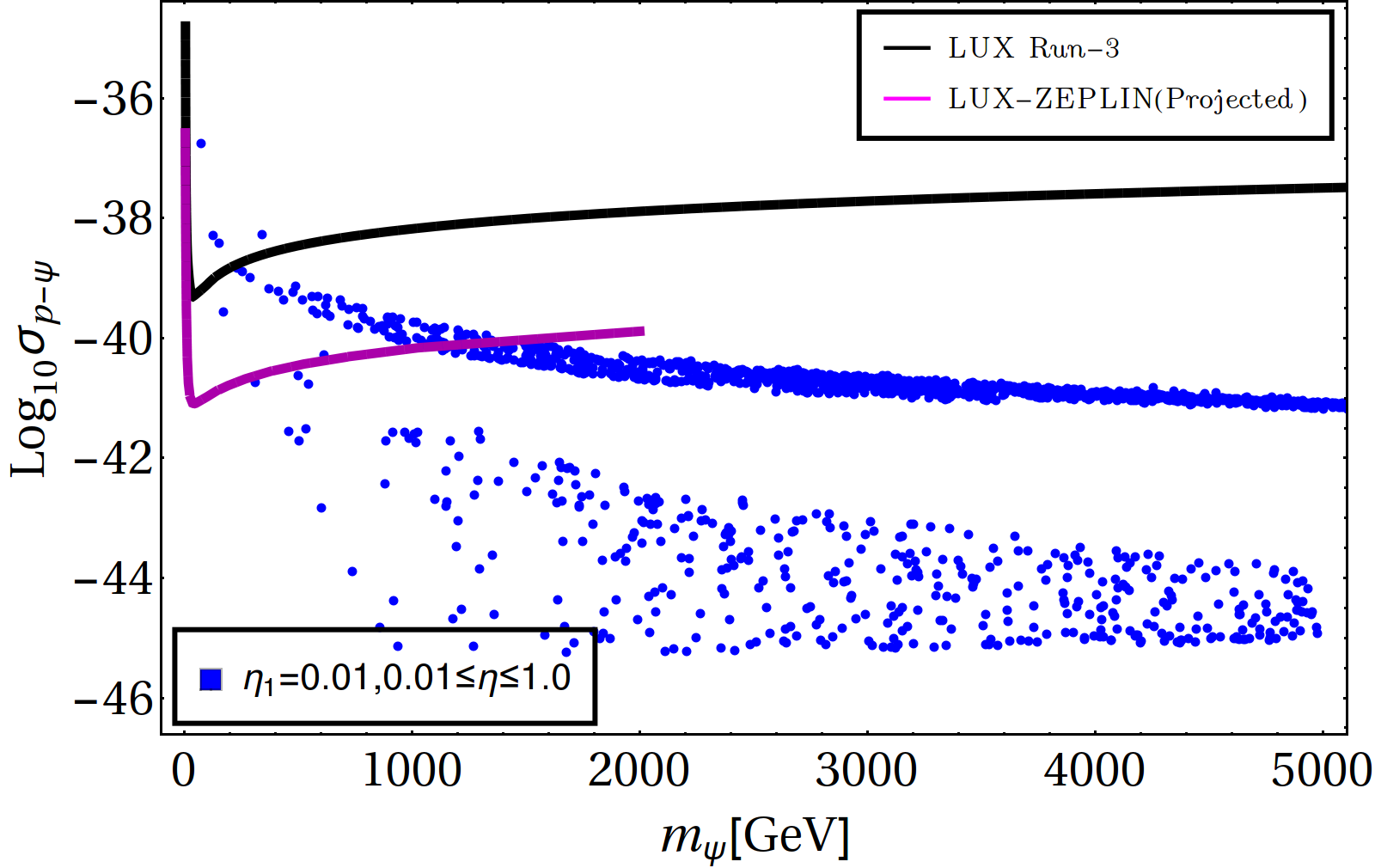}
$$
\caption{Left:DM-neutron SD scattering that satisfies PLANCK data on relic
  abundance where limits from LUX and proposed LUX-ZEPLIN have been shown.
  Right: Same but for DM-proton SD scattering.}
\label{fig:nulimits}
\end{figure}

The situation can be improved a lot when one assumes a different value of the coupling parameter for DM-torsion ($\eta$) relative to the SM fermions-torsion which is parameterized by $\eta_1$, as introduced in
Eq.~\ref{action_comp}. As discussed, we refer this scenario as non-universal torsion matter coupling (NU) model. In this case, we always keep $\eta_1 \sim 0.01$ which allows $m_s>3$~TeV to comply with the LHC dilepton results, while $\eta$ has been varied within $\{0.1-1\}$. With this choice of torsion-matter couplings, $Br(S_\mu \to l^+l^-)$ would be further reduced, thanks to rise in invisible branching fraction of the torsion. In the left side of Fig.~\ref{fig:nulimits} we place bounds on $m_s$ for
different choices of $\eta=0.05,0.5,1$ keeping $\eta_1=0.01$. For example for $\eta=0.05$, the
bounds on $m_s$ reduced to $m_s>1.4$~TeV from dilepton searches. Further rise in $\eta$ would make torsion mass even less constraint due to increase in the
invisible branching fractions. Similar to dilepton
searches, dijet bounds would not be of much significance. The relevant limits may come from the monojet searches which is induced by the axial vector couplings~\cite{Aaboud:2017phn,Sirunyan:2017hci}. According to an analysis in~\cite{Sirunyan:2017hci}, which corresponds to 13 TeV CM energy
and 12.9 $fb^{-1}$ integrated luminosity, torsion mass upto $\sim$ 2 TeV has been ruled out if one assumes a coupling strength of 0.25 between the mediators and the SM fermions while a coupling strength of 1.0 between the mediators and the DM particles. Similar limit $m_s > 1.55$TeV can be set
corresponding to an integrated luminosity of 36.1$fb^{-1}$~\cite{Aaboud:2017phn}. For our chosen input parameters $\eta_1$ and $\eta$, the limits should get much less stringent due to strong suppression in the production cross-section for $S_\mu$. In the RHS of Fig.\ref{fig:nulimits}, we depict the allowed
parameter points by PLANCK data. Acceptable relic abundance can be obtained along two regions: (i) resonance annihilations, which pass almost through the middle of the plane and (ii) $s$ and $t$-channel processes when the torsion is lighter than the DM. A detailed study of mono-jet searches on the parameter space may further constrain the parameter space. However, this is beyond the scope of this paper.



Finally, we also explore the parameter space via the SD direct detections in the plane of neutron-DM and proton-DM plane. Like our previous results in the universal coupling scenario, unpaired neutrons in Xenon based detectors would help to make neutron-DM scattering more effective to constrain the parameter space. Part of the parameter space, specially the resonance annihilation regions, can be probed in the near future via LZ experiments.

\section{Conclusions}
\label{sec:concl}

In this paper we have explored the possibility of having a fermionic dark matter (DM) in an extension of the SM, where torsion field ($S_\mu$) plays the role of portal. Guided by its geometric origin, the torsion field can naturally be realized as a massive axial vector field. The dark sector is comprised only of the SM-singlet Dirac fermion ($\psi$) that makes up all of the observed DM relic abundance. The DM is naturally stable over the cosmological time scale ({\it i.e.,} non-decaying) without the imposition of any ad-hoc symmetry or kinematical condition to forbid its decay (in other words, the DM need not to be the lightest stable particle (LSP) in this set-up). Thus the proposed model is a minimal extension of the SM. We consider two scenarios to parametrize the torsion-matter (includes both DM and SM particles) couplings. In the first case both the torsion-DM and torsion-SM couplings are assumed to be equal (dubbed as universal coupling scenario) at the low energy scale. It leads to a simplified DM phenomenology involving only three free parameters: the universal torsion-DM coupling $\eta$, DM mass $m_{\psi}$ and the torsion mass $m_s$. We consider the DM to be WIMP-like, and via a detailed parameter space scan we find that correct relic abundance can be achieved over a large parameter space for DM mass $m_{\psi}:\{1-5000\}~\rm GeV$, with torsion mass lying in the same range and for $\eta\sim\{0.1-1\}$. The DM phenomenology is somewhat equivalent to popular $Z^{'}$-mediated DM models but unlike $Z^{'}$, the torsion can only accommodate axial couplings. However, a significant difference lies in the fact that in this set-up the dark sector and the visible sector fermions can naturally enjoy the same coupling with the mediator. Parameter space of such a simple model is highly constrained from the recent LHC data.  Thus, stringent bounds can be derived from the dilepton searches at the LHC similar to massive (axial) $Z^{'}$ searches. From our computation of the observable $\sigma(pp \to S_\mu)\times Br(S_\mu \to l^+l^-)$ ($l\in e,\mu$), we find that $\eta \lsim 0.01 (0.1)$ is required for $m_s \lsim 3 (5)$~TeV which places very tight constraint on the viable DM parameter space. In fact, the whole parameter space is mostly excluded except for $\eta \sim 1$, where large decay width of the torsion may invalidate the LHC limits. Then we move to the non universal coupling scenario where torsion-DM interaction strength is assumed to be different from that of torsion-SM fermion (dubbed as non-universal coupling). Since LHC constraints are mainly determined by torsion-quark couplings, we set them at 0.01 and vary the DM-torsion couplings in the range \{0.1-1\}. Here $Br(S_\mu\to l^+l^-)$ is reduced due to rise in the invisible decay of the torsion, which essentially helps to have even lighter $S_\mu < 3$~TeV compared to the universal scenario. The PLANCK-observed relic abundance is satisfied along the resonance region and also for $m_s \le m_\psi$. In both universal and non-universal scenarios direct detection of the DM has been computed. Here, the presence of the axial-vector interaction turns out to be beneficial as this gives rise only to a spin-dependent direct detection cross-section. We find that almost all of the relic density allowed parameter space lies comfortably below the present exclusion limit of direct search experiments. However, future spin-dependent direct search experiments {\it e.g.,} LUX-ZEPLIN with higher sensitivity can probe some parts of the parameter space.

\section{Acknowledgments}
The computation was supported in part by the SAMKHYA: High Performance Computing Facility provided by Institute of Physics, Bhubaneswar. TB acknowledges the local hospitality at IOP, Bhubaneswar where this work was started. DD acknowledges
discussion with Sujoy Poddar. 


\bibliography{ref}
\end{document}